\documentclass[aps,prb,reprint,superscriptaddress,nofootinbib,floatfix]{revtex4-2}

\usepackage[utf8]{inputenc}
\usepackage[T1]{fontenc}
\usepackage{newtxtext,newtxmath}
\usepackage{amsmath,bm}
\usepackage{graphicx}
\usepackage{epstopdf}
\usepackage{booktabs}
\usepackage{array}
\usepackage{makecell}
\usepackage{pifont}
\newcommand{\cmark}{\ding{51}}
\usepackage[colorlinks=true,linkcolor=blue,citecolor=blue,urlcolor=blue]{hyperref}

\usepackage{color}

\begin{document}

\title{Piezoaxial coupling for strain-selected ferroaxial domain control}

\author{Rikuto Oiwa and Satoru Hayami}
\affiliation{Graduate School of Science, Hokkaido University, Sapporo 060-0810, Japan}

%%%%%%%%%%%%%%%%%%%%%%%%%%%%%%%%%%%%%%%%%%%%%%%%%%
\begin{abstract}
We formulate a symmetry-based hierarchy of strain-derived conjugate fields for ferroaxial order, and demonstrate strain-selected ferroaxial domain control using first-principles calculations.  
Since ferroaxial order is even under both spatial inversion and time reversal, ordinary electric and magnetic fields cannot serve as universal linear conjugate fields.
Homogeneous strain, however, can generate symmetry-allowed piezoaxial fields whose leading order is determined by the parent point group and by the chosen ferroaxial-axis component.  
For basal-plane strain, the leading field is linear in orthorhombic systems, quadratic in tetragonal systems, and cubic in trigonal and hexagonal systems. 
Cubic parent groups further split into two classes: cubic-I groups, $23$ and $m\bar{3}$, allow linear full-strain fields for selected axes, whereas cubic-II groups, $432$, $\bar{4}3m$, and $m\bar{3}m$, forbid linear fields and require quadratic or cubic strain combinations depending on the selected axis.
In trigonal systems, the basal-plane deviatoric strain with signed amplitude $\varepsilon_{\rm u}$ and principal-axis angle $\theta$ gives the single-axis field $h\propto\varepsilon_{\rm u}^3\sin6\theta$.
First-principles calculations for the trigonal ferroaxial compound Na$_2$BaMg(PO$_4$)$_2$ verify both the predicted angular dependence and cubic strain scaling of the ferroaxial domain splitting, and fixed-strain atomic relaxations show strain-selected evolution from the para-axial structure.
These results establish static homogeneous strain as a symmetry-allowed conjugate field for ferroaxial order and suggest a route to ferroaxial domain control through strain-field cooling.
\end{abstract}

\maketitle

%%%%%%%%%%%%%%%%%%%%%%%%%%%%%%%%%%%%%%%%%%%%%%%%%%
\section{Introduction}

Ferroaxial order is a ferroic order whose order parameter transforms as an axial vector and is even under spatial inversion $\mathcal{P}$ and time reversal $\mathcal{T}$~\cite{Hlinka2014,Hlinka2016}.  
It is therefore symmetry-distinct from $\mathcal{P}$-odd ferroelectric order and $\mathcal{T}$-odd ferromagnetic order.  
Equivalently, in the language of symmetry-adapted multipoles, ferroaxial order is represented by an electric-toroidal dipole $\bm{G}$~\cite{JPSJ.87.033709,PhysRevB.98.165110,KusunoseHayami2022JPCM,SH_JPSJ_review_2024}. 
Throughout this paper, we consider a selected single-axis ferroaxial order parameter, denoted by $A$, which, when necessary, is understood as the local axial-vector component $A_z^{\rm loc}$ defined with respect to a chosen local $z$ axis. 
Since ordinary electric and magnetic fields do not possess the same $\mathcal{P}$-even, $\mathcal{T}$-even axial-vector symmetry, neither can serve as a universal linear conjugate field for controlling the sign of a ferroaxial domain.

Ferroaxial order has been discussed in a broad range of materials.  
Representative structural displacive-type examples, in which $A$ can be associated with a collective rotation of structural units such as oxygen polyhedra, include the glaserite-related molybdate RbFe(MoO$_4$)$_2$~\cite{Jin_RbFeMoO42_2020,Owen2021PRB,Hayashida2021PRM,Zeng2025Science}, glaserite-type phosphates such as K$_2$Zr(PO$_4$)$_2$~\cite{Yamagishi2023Glaserite,Bhowal2024PRR,Xie2026PRB} and Na$_2$Ba$M$(PO$_4$)$_2$ ($M={\rm Mg},{\rm Mn},{\rm Co},{\rm Ni}$)~\cite{Kajita2024ChemMater,Kajita2025JPSJ}, Na$_2$Hf(BO$_3$)$_2$~\cite{Nagai2023Na2HfBO3}, and NASICON-type compounds~\cite{Nagai2023NASICON}.
Ferroaxial domains have also been directly visualized in order--disorder-type ferroaxial crystals such as NiTiO$_3$~\cite{Hayashida2020NatCommun,Hayashida2021PRM,Yokota2022npj,Guo2023PRB,Kusuno2026PRL}.
Beyond structural rotations of polyhedra, ferroaxial moment has also been discussed in electronic systems, including the hidden-order phase of URu$_2$Si$_2$~\cite{PhysRevB.101.205114, Hayami_URu2Si2_2023} and Ca$_5$Ir$_3$O$_{12}$~\cite{Hasegawa_ferro_rotation_2020,Hanate2021Ca5Ir3O12Superlattice,Hanate2023Ca5Ir3O12SpaceGroup,Hayami2023Ca5Ir3O12Cluster}.
More recently, ferroaxial charge-density-wave (CDW) states have %also 
been discussed in 1T-TiSe$_2$~\cite{Fu2016TiSe2Strain,Jiang2026TiSe2} and the rare-earth tritellurides $R$Te$_3$~\cite{Singh2025RTe3}.
These examples show that ferroaxial order can appear either as a structural rotation pattern of polyhedra or as an electronic composite or CDW instability.

\begin{table*}[t!]
\caption{
Representative probes, response fingerprints, and control routes for the single-axis ferroaxial order $A$.  
Here, ``Probe'' denotes a relatively direct readout of ferroaxial domains, ``Fingerprint'' denotes a response tensor or cross-correlated response enabled by ferroaxial order, and ``Control'' denotes an externally tunable field or effective field that can bias ferroaxial domain populations or switch ferroaxial domains.
The present work concerns the last row: a static homogeneous strain-derived axial field $h(\bm{\varepsilon})$, which couples to the ferroaxial order through the equilibrium free-energy term $-h(\bm{\varepsilon})A$ in Eq.~(\ref{eq:landau}) and biases domain populations during strain-cooling rather than switching already formed ferroaxial domains.
}
\label{tab:control}
\centering
\footnotesize
\renewcommand{\arraystretch}{1.2}
\setlength{\tabcolsep}{4pt}
\resizebox{\textwidth}{!}{%
\begin{tabular}{@{}lll@{}}
\toprule
Route & Representative observable or effective field & Type \\
\midrule
Linear electrogyration~\cite{Hayashida2020NatCommun,Yamagishi2023Glaserite,Kajita2025JPSJ}
&
\makecell[l]{Electric-field-induced optical rotation along the ferroaxial axis;\\$\Delta \theta_{z} \propto \gamma_{zzz} E_{z}$, with $\gamma_{zzz} \propto A$}
& 
Probe  \\
\addlinespace[0.35em]
Second-harmonic generation (SHG)~\cite{Jin_RbFeMoO42_2020,Owen2021PRB,Sekine2024MnTiO3}
& \makecell[l]{Electric-quadrupole (EQ) rotational anisotropy SHG response; \\ $I^{2\omega}_{\parallel,D_\pm}(\varphi) \propto \left(\chi^{\rm EQ}_{yyzy}\cos3\varphi \pm \chi^{\rm EQ}_{xxzx}\sin3\varphi\right)^2$, with $\chi^{\rm EQ}_{xxzx} \propto A$}
& Probe \\
\addlinespace[0.35em]
Raman optical activity~\cite{Kusuno2026PRL,Watanabe2025PRB,Suganuma2026Pyrite,Watanabe2026DualCircular}
& \makecell[l]{Circular-polarization-dependent Raman intensity;\\$g_{\rm ROA} = 2(I_{\rm LR}-I_{\rm RL}) / (I_{\rm LR}+I_{\rm RL}) \propto A$}
& Probe \\
\addlinespace[0.35em]
Superlattice reflections~\cite{Hanate2021Ca5Ir3O12Superlattice,Hanate2023Ca5Ir3O12SpaceGroup,Hayami2023Ca5Ir3O12Cluster}
& \makecell[l]{Superlattice peaks at $\bm{q} = (1/3,1/3,1/3)$ in Ca$_5$Ir$_3$O$_{12}$;\\space-group analysis identifies a finite-$q$ %hidden 
ferroaxial order}
& Probe \\
\hline
\addlinespace[0.35em]
\makecell[l]{Piezoresistivity~\cite{DayRoberts2025PRL} / elastoresistivity~\cite{Jiang2026TiSe2}}
& \makecell[l]{Symmetry-adapted off-diagonal strain-resistivity responses;\\$\partial\rho_{x^2-y^2}/\partial Y - \partial\rho_{2xy}/\partial X\propto A$ \\
Here, $\rho_{ij}$ is the resistivity tensor, where $\rho_{x^2-y^2} \equiv \rho_{xx} - \rho_{yy}$ and $\rho_{2xy} \equiv 2\rho_{xy}$. 
}
& Fingerprint \\
\addlinespace[0.35em]
Thermal cross-correlation~\cite{Nasu2022PRB}
& \makecell[l]{Antisymmetric thermopolarization;\\$P_x\propto\beta_{xy}(-\nabla_yT)$, with $\beta_{xy}=-\beta_{yx}\propto A$}
& Fingerprint \\
\addlinespace[0.35em]
Longitudinal spin-current~\cite{Roy2022SpinHall,Hayami2022ElectricFerroAxial}
& \makecell[l]{Spin-current parallel to the electric field;\\$J_{\mu}^{s_z} = \sigma^{z({\rm s})}_{\mu\mu} E_{\mu}$, with $\sigma^{z({\rm s})}_{xx} = \sigma^{z({\rm s})}_{yy} \propto A$}
& Fingerprint \\
\addlinespace[0.35em]
Planar Hall and magnetoresistive responses~\cite{Hayami2023PRB}
& \makecell[l]{Unconventional planar Hall and magnetoconductivity tensors;\\$\sigma^{\rm H}_{yz;y} = -\sigma^{\rm H}_{zx;x} \propto A$, $\sigma^{\rm MC}_{xy;xx} = -\sigma^{\rm MC}_{xy;yy} \propto A$}
& Fingerprint \\
\addlinespace[0.35em]
Nonlinear magnetoelastic response~\cite{Kirikoshi2023JPSJ}
& \makecell[l]{Second-order magnetic-field-induced strain; e.g.\\$X = \chi_{X, xy} H_{x} H_{y}$, with $\chi_{X,xy} \propto A$ \\
Here, $X\equiv\varepsilon_{xx}-\varepsilon_{yy}$.
}
& Fingerprint \\
\addlinespace[0.35em]
Nonlinear transverse magnetic response~\cite{Inda2023JPSJ,Du2026NatPhys}
& \makecell[l]{Third-order transverse magnetization; e.g.\\$M_{x} = \chi_{xyyy} H_{y}^3$, with $\chi_{xyyy} = -\chi_{yxxx} \propto A$}
& Fingerprint \\
\addlinespace[0.35em]
Nonlinear light-induced Edelstein effect~\cite{Kirikoshi2026NLEE}
&
\makecell[l]{Light-induced static magnetization from a second-order optical response;\\
$\delta M_\mu^{\rm LPL} = \sum_{\nu\lambda} \eta^{\rm spin}_{\mu;\nu\lambda} L_{\nu\lambda}$, \, $\delta M_\mu^{\rm CPL} = \sum_{\rho} \xi^{\rm spin}_{\mu;\rho} F_\rho$, with $\eta^{\rm spin}_{y;yz},\,\xi^{\rm spin}_{y;x} \propto A$. \\
Here, LPL and CPL denote linearly and circularly polarized light, respectively; \\
$L_{\nu\lambda}={\rm Re}[E_{\nu}(\Omega)E_{\lambda}^{\ast}(\Omega)]$ and $F_\rho=-(i/2)[\bm{E}(\Omega)\times\bm{E}^{\ast}(\Omega)]_\rho$.
}
& Fingerprint \\
\hline
\addlinespace[0.35em]
Light-induced circular phonons~\cite{HeKhalsa2024PRR,Zeng2025Science}
& \makecell[l]{Dynamical axial field from light-induced circular phonons;\\$-h(t) A$, $h(t) \propto (\bm{Q}_{\rm IR} \times \bm{E}_{\rm THz})_{z}$ \\
Here, $\bm{Q}_{\rm IR}$ is the displacement vector of a resonantly driven infrared-active phonon \\
and $\bm{E}_{\rm THz}$ is the electric field vector of the circularly polarized terahertz pulse.
}
& Control \\
\addlinespace[0.35em]
Homogeneous strain (this work; cf.~\cite{Jiang2026TiSe2})
& \makecell[l]{Static strain polynomial $h(\bm{\varepsilon})$;\\$-h(\bm{\varepsilon})A$, e.g. $h \propto Y(3X^2-Y^2)$ for trigonal basal-plane strain \\
Here, $X\equiv\varepsilon_{xx}-\varepsilon_{yy}$, $Y\equiv2\varepsilon_{xy}$.
}
& Control \\
\bottomrule
\end{tabular}%
}
\end{table*}

Existing experimental and theoretical studies can be broadly classified into three categories, as summarized in Table~\ref{tab:control}.  
The first category is direct or near-direct probing of ferroaxial domains.  
Linear electrogyration images ferroaxial domains through an electric-field-induced optical rotation whose sign follows the ferroaxial order parameter $A$~\cite{Hayashida2020NatCommun,Yamagishi2023Glaserite,Kajita2025JPSJ}.  
Electric-quadrupole second-harmonic generation (SHG) distinguishes mirror-related ferroaxial domains through their rotational-anisotropy patterns~\cite{Jin_RbFeMoO42_2020,Owen2021PRB}, and related SHG imaging has been used for MnTiO$_3$~\cite{Sekine2024MnTiO3}.  
Raman optical activity~\cite{Kusuno2026PRL,Watanabe2025PRB,Suganuma2026Pyrite,Watanabe2026DualCircular} and superlattice-reflection measurements~\cite{Hanate2021Ca5Ir3O12Superlattice,Hanate2023Ca5Ir3O12SpaceGroup,Hayami2023Ca5Ir3O12Cluster} provide additional optical and diffraction routes to ferroaxial orders.

The second category is response fingerprints.  
Examples include piezoresistivity~\cite{DayRoberts2025PRL}, elastoresistivity~\cite{Jiang2026TiSe2}, antisymmetric thermopolarization~\cite{Nasu2022PRB}, longitudinal spin-current generation~\cite{Roy2022SpinHall,Hayami2022ElectricFerroAxial}, planar Hall and magnetoresistive responses~\cite{Hayami2023PRB}, nonlinear magnetoelasticity~\cite{Kirikoshi2023JPSJ}, nonlinear transverse magnetization~\cite{Inda2023JPSJ,Du2026NatPhys}, and nonlinear light-induced Edelstein effect~\cite{Kirikoshi2026NLEE}.
In these phenomena, symmetry allows tensor components that are proportional to $A$ and therefore reverse sign between opposite ferroaxial domains, providing characteristic fingerprints of ferroaxial order.

The third category is control.  
A conjugate field for ferroaxial order $h$ must itself be a $\mathcal{P}$-even, $\mathcal{T}$-even axial quantity so that the free energy can contain a term $-hA$.  
A dynamical route has recently been established using circularly polarized terahertz pulses~\cite{HeKhalsa2024PRR,Zeng2025Science}.  
When an infrared-active phonon is resonantly driven, the phonon displacement $\bm{Q}_{\rm IR}$ and the terahertz electric field $\bm{E}_{\rm THz}$ form the composite axial field $(\bm{Q}_{\rm IR}\times\bm{E}_{\rm THz})_z$, whose sign is controlled by the light helicity.  
This mechanism provides a powerful route to ferroaxial switching, but it relies on a nonequilibrium, dynamically generated axial field.  
The question addressed here is whether a static homogeneous strain field can play an analogous role in equilibrium.

The search for a static conjugate field is partly motivated by the recently proposed piezochiral effect~\cite{Zeng2025Piezochiral}, which addresses an analogous symmetry problem for chiral order.  
Chirality, like ferroaxiality, lacks a universal simple external conjugate field.  
A chiral order parameter is a $\mathcal{P}$-odd, $\mathcal{T}$-even pseudoscalar: it changes sign under spatial inversion or mirror operations while remaining invariant under proper rotations and $\mathcal{T}$ operation~~\cite{L.D.Barron_1986_true-chirality,Barron_mol-light-scattering}.  
By contrast, the ferroaxial order considered here is a $\mathcal{P}$-even and $\mathcal{T}$-even axial-vector component.  
Although homogeneous strain $\varepsilon_{ij}$ is a $\mathcal{P}$-even, $\mathcal{T}$-even symmetric tensor, $\varepsilon_{ij} = \varepsilon_{ji}$, the symmetry channels through which it couples to chirality and ferroaxiality are fundamentally different.
In the piezochiral effect, suitable strain components or strain polynomials transform as a pseudoscalar field and can induce or control handedness in an achiral crystal, with the sign determined by the strain direction or by switching between tensile and compressive strain. 
The ferroaxial case is analogous in spirit but distinct in symmetry: the relevant order parameter is not a pseudoscalar but a $\mathcal{P}$-even, $\mathcal{T}$-even axial vector, transforming in the same point-group representation as $A$.

Experimental signatures of strain--ferroaxial coupling have recently emerged in ferroaxial systems~\cite{DayRoberts2025PRL,Jiang2026TiSe2}.
In particular, elastoresistivity measurements on 1T-TiSe$_2$ showed that a cubic combination of in-plane deviatoric strains acts as an effective conjugate field to the ferroaxial CDW order and exhibits hysteresis associated with ferroaxial domain-wall motion~\cite{Jiang2026TiSe2}.  
These results provide direct evidence that strain can manipulate ferroaxial order.
However, a general symmetry framework for identifying such strain-derived conjugate fields across all point groups, together with first-principles verification of the resulting ferroaxial energy splitting, has yet to be established.

Motivated by these developments, we formulate a symmetry-based hierarchy of the lowest-order static homogeneous strain polynomials that transform in the same irreducible representation as the ferroaxial order parameter and therefore act as an effective conjugate field $h(\bm{\varepsilon})$. 
We then verify the trigonal case by first-principles total-energy calculations for the displacive-type ferroaxial compound Na$_2$BaMg(PO$_4$)$_2$.  
This establishes the concept of piezoaxial coupling: a static homogeneous strain generates an equilibrium axial field that energetically selects the ferroaxial domain.

\begin{figure*}[t!]
\includegraphics[width=0.98\linewidth]{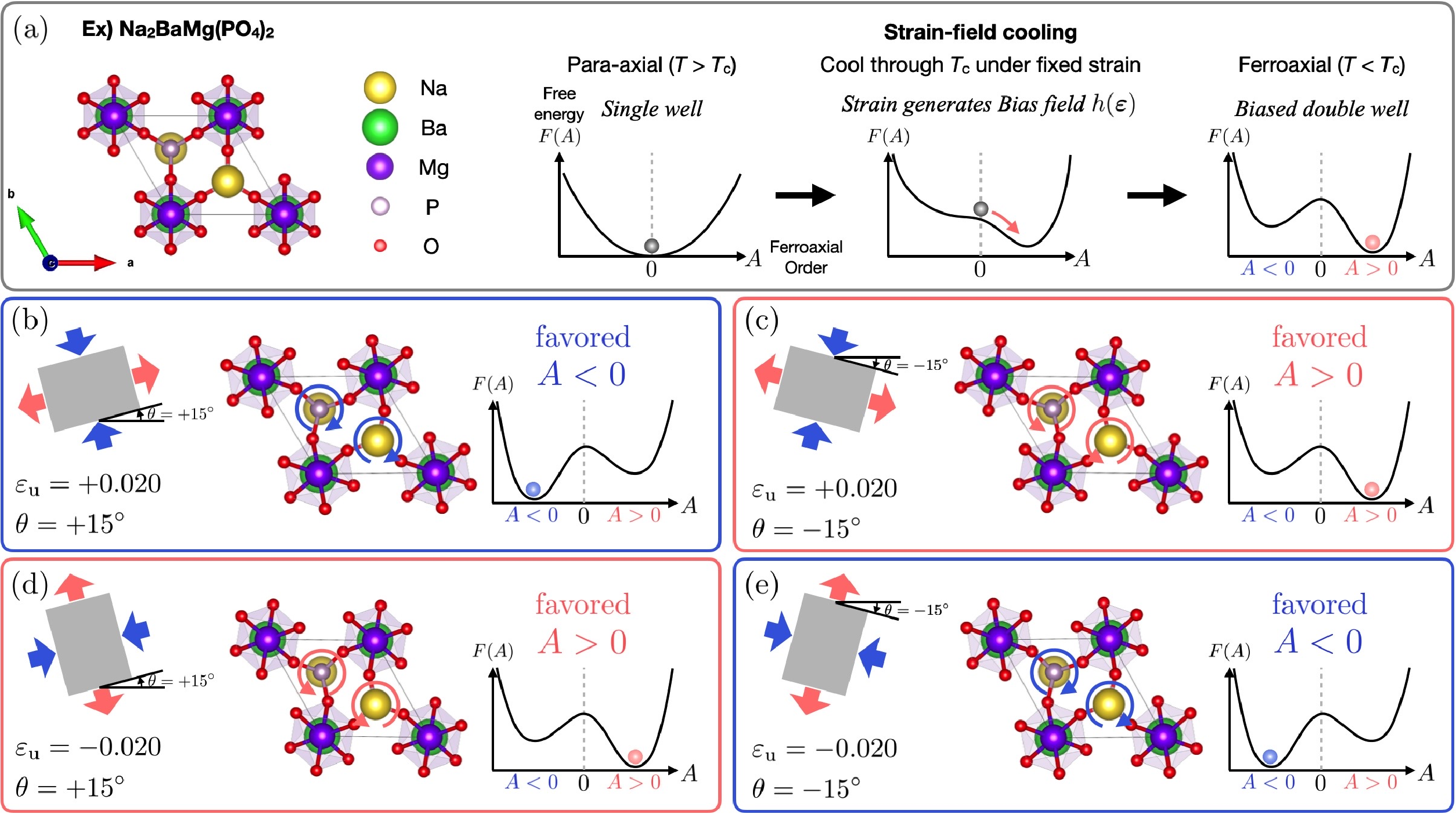}
\caption{
Strain-field cooling protocol for selecting ferroaxial domains.
(a) Schematic illustration of domain selection in glaserite-type ferroaxial compound Na$_2$BaMg(PO$_4$)$_2$.  
Above the ferroaxial transition temperature, $T_{\rm c}\simeq540$ K~\cite{Kajita2024ChemMater}, the para-axial phase has a single free-energy minimum at $A = 0$.
A fixed static homogeneous strain generates a symmetry-allowed conjugate field $h(\bm{\varepsilon})$, which couples linearly to the ferroaxial order parameter and lifts the degeneracy of the ferroaxial double-well potential upon cooling through $T_{\rm c}$, thereby selecting one of the two ferroaxial domains. 
(b)--(e) Four representative strain-field-cooling conditions for $\varepsilon_{\rm u} = \pm 0.020$ and $\theta = \pm 15^\circ$.
The gray parallelogram and colored arrows schematically indicate the applied strain, while the colored circular arrows indicate the selected sense of the local rotations, $A > 0$ (red) or $A < 0$ (blue).
The corresponding biased double-well potentials show the favored ferroaxial domain.  
For the convention used here, the strain-induced bias field is proportional to $\varepsilon_{\rm u}^{3} \sin 6\theta$; therefore, reversing either the sign of $\varepsilon_{\rm u}$ or the sign of $\theta$ reverses the favored ferroaxial domain.  
Specifically, $(\varepsilon_{\rm u},\theta) = (+0.020,+15^\circ)$ and $(-0.020,-15^\circ)$ favor $A < 0$, whereas $(+0.020,-15^\circ)$ and $(-0.020,+15^\circ)$ favor $A > 0$.
}
\label{fig:concept}
\end{figure*}

Figure~\ref{fig:concept} schematically illustrates the central concept of this work: a fixed static homogeneous strain generates symmetry-allowed axial field $h(\bm{\varepsilon})$ that couples linearly to the ferroaxial order parameter $A$ through $F_{\rm A} = -h(\bm{\varepsilon})A$.
Cooling through the ferroaxial transition under this strain-derived field lifts the degeneracy between the two ferroaxial domains and biases their population.
We refer to this symmetry-allowed invariant as the piezoaxial coupling.
As a representative example, we consider the trigonal ferroaxial compound Na$_2$BaMg(PO$_4$)$_2$.
For this symmetry, the leading basal-plane contribution is cubic in the signed deviatoric strain amplitude, $h(\bm{\varepsilon})\propto \varepsilon_{\rm u}^{3}\sin6\theta$, where $\varepsilon_{\rm u}$ denotes the signed magnitude of the traceless basal-plane strain and $\theta$ is its principal-axis angle.
Thus, reversing the sign of $\varepsilon_{\rm u}$ or rotating the strain principal axis by $30^\circ$ reverses the axial field and selects the opposite ferroaxial domain during strain cooling.

The rest of the paper develops this concept quantitatively.
Section~\ref{sec:formulation} formulates the strain-derived axial field and classifies the lowest-order strain polynomials allowed by point-group symmetry.
Section~\ref{sec:demo} presents the first-principles demonstration for Na$_2$BaMg(PO$_4$)$_2$: the calculated clamped-coordinate domain splitting follows the predicted $\sin6\theta$ angular dependence and $\varepsilon_{\rm u}^{3}$ scaling, the strained double-well potential exhibits the corresponding domain bias, and fixed-strain relaxations from the para-axial structure provide an additional check of strain-selected basin preference.
We also estimate the possible relevance of the calculated bias for strain-cooling experiments.
Section~\ref{sec:scope} discusses the scope and limitations of the proposal and its relation to other material classes.
Section~\ref{sec:conclusion} summarizes the conclusions.

%%%%%%%%%%%%%%%%%%%%%%%%%%%%%%%%%%%%%%%%%%%%%%%%%%
\section{General formulation}
\label{sec:formulation}

We first formulate the static homogeneous strain-derived conjugate field for ferroaxial order.
Let $\bm{A}$ be the ferroaxial axial-vector order parameter, and let $\hat{\bm n}$ denote the ferroaxial axis component to be biased.  
The selected component is
\begin{equation}
 A_{\hat{\bm{n}}} = \hat{\bm{n}} \cdot \bm{A}.
 \label{eq:selected_component}
\end{equation}
We then introduce a local orthonormal frame $(\hat{\bm{x}}_{\rm loc},\hat{\bm{y}}_{\rm loc},\hat{\bm{z}}_{\rm loc})$ whose local $z$ axis is chosen as $\hat{\bm z}_{\rm loc} = \hat{\bm n}$.  
In this local frame, the selected component is denoted by
$A_z^{\rm loc}$, or simply by $A$, when no ambiguity arise.  
This convention does not imply that the local $z$ axis coincides with the conventional crystallographic $z$ axis.  
For non-cubic parent groups, $\hat{\bm{n}}$ is usually fixed by the crystallographic $z$ axis.
For cubic parent groups, the full axial vector $\bm{A}$ must first be treated under the parent cubic symmetry, and the corresponding component, such as $A_{[001]}$, $A_{[110]}$, and $A_{[111]}$, is obtained by projecting $\bm{A}$ onto the chosen cubic direction.

%%%%%%%%%%
\subsection{Piezoaxial coupling}

Let $\bm{\varepsilon}$ denote the homogeneous strain tensor.  
Its six independent components are written as $\varepsilon_{\mu}$ in Voigt notation, with $\mu=1,\ldots,6$ corresponding to $xx$, $yy$, $zz$, $yz$, $zx$, and $xy$, respectively.  
The central object of this work is a static homogeneous strain-derived axial field
\begin{equation}
 h(\bm{\varepsilon}),
 \label{eq:h_definition}
\end{equation}
which transforms in the same irreducible representation as the selected ferroaxial component.  
Its form is entirely determined by the parent point-group symmetry.  
Although its overall magnitude depends on the normalization and microscopic definition of $A$, its transformation property, lowest allowed order in strain, and angular dependence are fixed by symmetry.

The most general polynomial expansion of $h(\bm{\varepsilon})$ is
\begin{align}
h(\bm{\varepsilon}) &= \sum_{n=1} P^{(n)}(\bm{\varepsilon}),
\label{eq:h_expansion}
\\
P^{(1)}(\bm{\varepsilon}) &= \sum_{\mu=1}^{6} g^{(1)}_{\mu}\varepsilon_{\mu},
\label{eq:poly1}
\\
P^{(2)}(\bm{\varepsilon}) &= \sum_{\mu,\nu=1}^{6} g^{(2)}_{\mu\nu}\varepsilon_{\mu}\varepsilon_{\nu},
\label{eq:poly2}
\\
P^{(3)}(\bm{\varepsilon}) &=  \sum_{\mu,\nu,\lambda=1}^{6} g^{(3)}_{\mu\nu\lambda} \varepsilon_{\mu}\varepsilon_{\nu}\varepsilon_{\lambda}.
\label{eq:poly3}
\end{align}
The coefficients $g^{(n)}$ are material dependent.
The parent point-group symmetry determines which tensor components can be nonzero, and hence whether the leading piezoaxial coupling is linear, quadratic, cubic, or higher order in strain.

The coupling to the ferroaxial order parameter, referred to here as \textit{piezoaxial coupling}, is written as
\begin{equation}
F_{\rm A} = -h(\bm{\varepsilon}) A .
\label{eq:bias_general}
\end{equation}
Here, $A$ denotes the selected axial-vector component in the local-frame convention defined above, and $h(\bm{\varepsilon})$ is the strain-derived field transforming in the same representation.  
Equation~(\ref{eq:bias_general}) defines the sign convention used below: a domain with $h(\bm{\varepsilon})A > 0$ is lowered in free energy.

The Landau free energy near the ferroaxial transition temperature $T_{\rm c}$ at fixed static homogeneous strain can be written as
\begin{multline}
F(A;\bm{\varepsilon}) = \frac{a}{2}(T-T_{\rm c})A^2 + \frac{b}{4}A^4 + \frac{c}{6}A^6 - h(\bm{\varepsilon})A
\\
+ E_{\rm el}(\bm{\varepsilon}) + F_{\rm mix}(A^2,\bm{\varepsilon}) + \cdots ,
\label{eq:landau}
\end{multline}
where $T$ is the temperature and $a$, $b$, and $c$ are Landau coefficients.
The sixth-order term is retained to allow for weakly first-order ferroaxial transitions~\cite{Yamagishi2023Glaserite}.
The term $E_{\rm el}(\bm{\varepsilon})$ is the ordinary elastic energy of the imposed strain, for example
\begin{equation}
E_{\rm el}(\bm{\varepsilon})
=
\frac{1}{2} \sum_{\mu,\nu=1}^{6} C_{\mu\nu}\varepsilon_{\mu}\varepsilon_{\nu} + O(\varepsilon^3),
\label{eq:elastic_energy}
\end{equation}
where $C_{\mu\nu}$ are the elastic constants.
The term $F_{\rm mix} (A^{2},\bm{\varepsilon})$ denotes strain couplings that are even in $A$, such as $A^2\varepsilon_{zz}$.
For a fixed imposed strain, %both 
$E_{\rm el}(\bm{\varepsilon})$, $F_{\rm mix}(A^2,\bm{\varepsilon})$, and the other even-in-$A$ terms, are identical for the two ferroaxial domains and therefore do not contribute to the domain-odd energy difference.

Below $T_{\rm c}$, the two ferroaxial domains are represented by the two order-parameter signs, $+A$ and $-A$ with $A > 0$.
To leading order in the strain-derived axial field, the domain splitting is
\begin{equation}
\Delta E_{\rm dom}(A; \bm{\varepsilon}) \equiv E(+A;\bm{\varepsilon}) - E(-A;\bm{\varepsilon}) = -2h(\bm{\varepsilon})A + O[h(\bm{\varepsilon})^3].
\label{eq:domain_split_landau}
\end{equation}
This domain splitting is the primary first-principles observable used in this paper.
Its sign determines which ferroaxial domain is thermodynamically favored under the imposed strain.

The coupling in Eq.~(\ref{eq:bias_general}) is formulated for a uniform macroscopic ferroaxial order parameter.
Accordingly, it is most naturally interpreted as a conjugate-field coupling for a $\Gamma$-point, or proper, ferroaxial order.
If the primary structural or electronic order parameter carries a finite wave vector $\bm{Q}$, a homogeneous strain field cannot couple linearly to that primary order parameter because of translational symmetry.
In such cases, the present coupling should instead be understood as a bias field for a uniform axial composite induced by the finite-$\bm{Q}$ order.
Schematically, if $\eta_{\bm{Q}}$ denotes the primary finite-wave-vector order parameter, the induced uniform ferroaxial component can arise as
\begin{equation}
A^{\rm ind} \sim \mathcal{I}_A[ \eta_{\bm{Q}},\eta_{-\bm{Q}}, \ldots ],
\label{eq:finite_q_composite}
\end{equation}
where the invariant $\mathcal{I}_A$ transforms as the selected axial-vector component and carries zero total wave vector.
The static homogeneous strain-derived field then biases $A^{\rm ind}$, and hence the corresponding ferroaxial domain, as experimentally shown in Ref.~\cite{Jiang2026TiSe2}, but it is not a direct conjugate field to $\eta_{\bm{Q}}$ itself.

%%%%%%%%%%
\subsection{Point-group hierarchy of piezoaxial strain fields}
\label{sec:point_group_hierarchy}

\begin{table*}[t!]
\caption{
Point-group hierarchy of strain-derived axial fields $h(\bm{\varepsilon})$ in non-pyroaxial parent point groups; details of the derivation are given in the Supplemental Material~\cite{SupplementalMaterial}.
The ferroaxial-order column specifies the axial component biased by the listed strain polynomial.
For non-cubic families, $A_z^{\mathrm{loc}}$ denotes the component along the symmetry-distinguished local $z$ axis, which need not coincide with a conventional crystallographic axis.
For cubic families, the order parameter is specified in the parent cubic frame as $A_{[001]}$, $A_{[110]}$, or $A_{[111]}$, and the subscript ``loc'' on strain variables denotes the local frame obtained by rotating the selected axis to the local $z$ axis.
The corresponding normalized local frames are $(\hat{\bm{x}}_{\mathrm{loc}},\hat{\bm{y}}_{\mathrm{loc}},\hat{\bm{z}}_{\mathrm{loc}}) =([100],[010],[001])$ for $[001]$, $([\bar{1}10]/\sqrt{2},[001],[110]/\sqrt{2})$ for $[110]$, and $([\bar{1}10]/\sqrt{2},[\bar{1}\bar{1}2]/\sqrt{6},[111]/\sqrt{3})$ for $[111]$.
The basal subgroup is the residual point group for a representative basal-plane strain with nonzero leading basal field, whereas the non-basal subgroup is that for a representative strain involving at least one of $Z$, $U$, and $V$.
Here $X=\varepsilon_{xx}-\varepsilon_{yy}$, $Y=2\varepsilon_{xy}$, $Z=2\varepsilon_{zz}-\varepsilon_{xx}-\varepsilon_{yy}$, $U=2\varepsilon_{xz}$, and $V=2\varepsilon_{yz}$ are defined in the selected local frame.
For a pure in-plane uniaxial deviatoric strain, $X=2\varepsilon_{\rm u}\cos2\theta$ and $Y=2\varepsilon_{\rm u}\sin2\theta$, giving the angular dependence shown in the leading-basal-field column.
}
\label{tab:classification}
\centering
\scriptsize
\renewcommand{\arraystretch}{1.18}
\setlength{\tabcolsep}{2.0pt}
\resizebox{\textwidth}{!}{%
\begin{tabular}{@{}lllllll@{}}
\toprule
\makecell[l]{Crystal\\family} & \makecell[l]{Parent point\\ group(s)} & \makecell[l]{Ferroaxial\\order} & \makecell[l]{Basal\\subgroup} & \makecell[l]{Leading basal\ field} & \makecell[l]{Non-basal\\subgroup} & \makecell[l]{Leading non-basal or\ full-strain field} \\
\midrule
Orthorhombic
& \makecell[l]{$222$, $mm2$\\$mmm$}
& $A_z^{\mathrm{loc}}$
& \makecell[l]{$2$\\$2/m$}
& \makecell[l]{1st: $Y=2\varepsilon_{xy}$\\$\propto\varepsilon_{\rm u}\sin2\theta$}
& \makecell[l]{$1$\\$\bar1$}
& 2nd: $YZ$, $UV$ \\
\addlinespace[0.35em]
Tetragonal
& \makecell[l]{$422$, $4mm$, $\bar4 2m$, $\bar4 m2$\\$4/mmm$}
& $A_z^{\mathrm{loc}}$
& \makecell[l]{$2$\\$2/m$}
& \makecell[l]{2nd: $XY=2(\varepsilon_{xx}-\varepsilon_{yy})\varepsilon_{xy}$\\$\propto\varepsilon_{\rm u}^2\sin4\theta$}
& \makecell[l]{$1$\\$\bar1$}
& 3rd: $XYZ$, $XUV$, $Y(V^2-U^2)$ \\
\addlinespace[0.35em]
Trigonal
& \makecell[l]{$32$\\$3m$\\$\bar3m$}
& $A_z^{\mathrm{loc}}$
& \makecell[l]{$1$\\$1$\\$\bar1$}
& \makecell[l]{3rd: $Y(3X^2-Y^2)$\\$\propto\varepsilon_{\rm u}^3\sin6\theta$}
& \makecell[l]{$1$\\$1$\\$\bar1$}
& \makecell[l]{2nd: $XU-YV$ for $32$, $\bar3m$\\2nd: $XV+YU$ for $3m$} \\
\addlinespace[0.35em]
Hexagonal
& \makecell[l]{$622$, $6mm$\\$\bar6m2$, $\bar6 2m$\\$6/mmm$}
& $A_z^{\mathrm{loc}}$
& \makecell[l]{$2$\\$m$\\$2/m$}
& \makecell[l]{3rd: $Y(3X^2-Y^2)$\\$\propto\varepsilon_{\rm u}^3\sin6\theta$}
& \makecell[l]{$1$\\$1$\\$\bar1$}
& 3rd: $2XUV-Y(U^2-V^2)$ \\
\midrule
Cubic I
& \makecell[l]{$23$\\$m\bar3$}
& $A_{[001]}$
& \makecell[l]{$2$\\$2/m$}
& \makecell[l]{1st: $Y_{\rm loc}=2\varepsilon_{xy}^{\rm loc}$\\$\propto\varepsilon_{\rm u}\sin2\theta$}
& \makecell[l]{$1$\\$\bar1$}
& 2nd: $Y_{\rm loc}Z_{\rm loc}$, $U_{\rm loc}V_{\rm loc}$ \\
\addlinespace[0.25em]
& \makecell[l]{$23$\\$m\bar3$}
& $A_{[110]}$
& \makecell[l]{$1$\\$\bar1$}
& \makecell[l]{2nd: $X_{\rm loc}Y_{\rm loc}$\\$\propto\varepsilon_{\rm u}^2\sin4\theta$}
& \makecell[l]{$1$\\$\bar1$}
& 1st: $V_{\rm loc}=2\varepsilon_{yz}^{\rm loc}$ \\
\addlinespace[0.25em]
& \makecell[l]{$23$\\$m\bar3$}
& $A_{[111]}$
& \makecell[l]{$1$\\$\bar1$}
& \makecell[l]{2nd: $X_{\rm loc}^2+Y_{\rm loc}^2$\\$\propto\varepsilon_{\rm u}^2$}
& \makecell[l]{$3$\\$\bar3$}
& 1st: $Z_{\rm loc}=2\varepsilon_{zz}^{\rm loc}-\varepsilon_{xx}^{\rm loc}-\varepsilon_{yy}^{\rm loc}$ \\
\addlinespace[0.35em]
Cubic II
& \makecell[l]{$432$\\ $\bar4 3m$, \\$m\bar3m$}
& $A_{[001]}$
& \makecell[l]{$2$\\$2$\\$2/m$}
& \makecell[l]{2nd: $X_{\rm loc}Y_{\rm loc}$\\$\propto\varepsilon_{\rm u}^2\sin4\theta$}
& \makecell[l]{$1$\\$1$\\$\bar1$}
& 3rd: $X_{\rm loc}Y_{\rm loc}Z_{\rm loc}$, $X_{\rm loc}U_{\rm loc}V_{\rm loc}$, $Y_{\rm loc}(V_{\rm loc}^2-U_{\rm loc}^2)$ \\
\addlinespace[0.25em]
& \makecell[l]{$432$\\ $\bar4 3m$, \\$m\bar3m$}
& $A_{[110]}$
& \makecell[l]{$2$\\$m$\\$2/m$}
& \makecell[l]{2nd: $X_{\rm loc}Y_{\rm loc}$\\$\propto\varepsilon_{\rm u}^2\sin4\theta$}
& \makecell[l]{$1$\\$1$\\$\bar1$}
& 2nd: $3X_{\rm loc}Y_{\rm loc}+Y_{\rm loc}Z_{\rm loc}+2U_{\rm loc}V_{\rm loc}$ \\
\addlinespace[0.25em]
& \makecell[l]{$432$\\ $\bar4 3m$, \\$m\bar3m$}
& $A_{[111]}$
& \makecell[l]{$1$\\$1$\\$\bar1$}
& \makecell[l]{3rd: $Y_{\rm loc}(3X_{\rm loc}^2-Y_{\rm loc}^2)$\\$\propto\varepsilon_{\rm u}^3\sin6\theta$}
& \makecell[l]{$1$\\$1$\\$\bar1$}
& \makecell[l]{2nd: $X_{\rm loc}U_{\rm loc}-Y_{\rm loc}V_{\rm loc}$} \\
\bottomrule
\end{tabular}%
}
\end{table*}

\begin{table*}[t!]
\caption{
Representative ferroaxial candidate systems corresponding to the symmetry classes in Table~\ref{tab:classification}.
The ferroaxial-order column specifies the axial order parameter considered for the strain-derived field.
For non-cubic entries, $A_z^{\mathrm{loc}}$ means that the coordinate frame is chosen so that the relevant ferroaxial axis is the local $z$ axis; this direction need not coincide with the conventional crystallographic $c$ axis and may correspond to a conventional $a$ or $b$ direction in a particular setting.
For cubic entries, $A_{[001]}$ or $A_{[111]}$ denotes the selected axis in the parent cubic frame; symmetry-equivalent axes are implied.
A check mark in the $\Gamma$-point column denotes no enlargement of the high-temperature unit cell.
Materials citing Ref.~\cite{Jain2013MaterialsProject} were first identified as same-composition non-pyroaxial/pyroaxial structural pairs in the Materials Project and retained only when supported by experimental structural data.
}
\label{tab:materials}
\centering
\scriptsize
\renewcommand{\arraystretch}{1.14}
\setlength{\tabcolsep}{2.0pt}
\resizebox{\textwidth}{!}{%
\begin{tabular}{@{}lllllll@{}}
\toprule
\makecell[l]{Symmetry\ class} & \makecell[l]{Representative\ system} & \makecell[l]{Ferroaxial\\order} & \makecell[l]{Low-$T$\\space group} & \makecell[l]{High-$T$\\space group} & \makecell[l]{$\Gamma$\ point} & \makecell[l]{Electronic\\structure} \\
\midrule
Orthorhombic & Ta$_2$NiSe$_5$~\cite{Watson2020Ta2NiSe5, Maraytta2025Ta2NiSe5} & $A_z^{\mathrm{loc}}$ & $C2/c$ & $Cmcm$ & \cmark & EI candidate \\
 & NbNiTe$_2$~\cite{Neu2019NbNiTe2} & $A_z^{\mathrm{loc}}$ & $P112_1/a$ & $Pmna$ & \cmark & metal \\
 & $R$VO$_3$ ($R={\rm Dy},{\rm Ho},{\rm Er}$)~\cite{Reehuis2011RVO3} & $A_z^{\mathrm{loc}}$ & $P2_1/b$ & $Pbnm$ & \cmark & Mott insulator \\
 & CeCu$_6$~\cite{Poudel2015CeCu6,Jain2013MaterialsProject} & $A_z^{\mathrm{loc}}$ & $P2_1/c$ & $Pnma$ & \cmark & heavy-fermion metal \\
 & LaTaO$_4$~\cite{Howieson2021LaTaO4,Jain2013MaterialsProject} & $A_z^{\mathrm{loc}}$ & $P2_1/c$ & $Cmc2_1$ &  & insulator \\
 & WO$_3^{\dagger}$~\cite{Howard2002WO3,Jain2013MaterialsProject} & $A_z^{\mathrm{loc}}$ & $P2_1/c$ & $Pbcn$ & \cmark & semiconductor/insulator \\
 & $R$NiO$_3$ ($R={\rm Pr},{\rm Nd},{\rm Sm},{\rm Ho},{\rm Y},{\rm Er},{\rm Lu}$)~\cite{GarciaMunoz1992RNiO3,Alonso2001RNiO3,Jain2013MaterialsProject} & $A_z^{\mathrm{loc}}$ & $P2_1/n$ & $Pbnm$ & \cmark & metal--insulator \\
 & $Ln_3$IrO$_7$ ($Ln={\rm Pr},{\rm Nd},{\rm Sm},{\rm Eu}$)~\cite{Nishimine2007Ln3IrO7,Jain2013MaterialsProject} & $A_z^{\mathrm{loc}}$ & $P2_1/n$ & $Cmcm$ &  & insulator \\
 & $R$Te$_3$ ($R={\rm La},{\rm Gd},{\rm Ho},{\rm Er}$)~\cite{Singh2025RTe3} & $A_z^{\mathrm{loc}}$ & $C_{2h}^{\ddagger}$ & $Cmcm$ &  & CDW metal \\
\addlinespace[0.25em]
Tetragonal & VO$_2$~\cite{Hlinka2016,Lee2016VO2Polymorphs,DelValle2025VO2} & $A_z^{\mathrm{loc}}$ & $P2_1/c$ & $P4_2/mnm$ &  & metal--insulator \\
% & URu$_2$Si$_2$~\cite{PhysRevB.101.205114, Hayami_URu2Si2_2023} & $A_z^{\mathrm{loc}}$ & $P4/nnc$ & $I4/mmm$ &  & heavy-fermion HO \\
\addlinespace[0.25em]
Trigonal & K$_2$Zr(PO$_4$)$_2$~\cite{Yamagishi2023Glaserite,Bhowal2024PRR,Xie2026PRB} & $A_z^{\mathrm{loc}}$ & $P\bar3$ & $P\bar3m1$ & \cmark & insulator \\
 & Na$_2$Ba$M$(PO$_4$)$_2$ ($M={\rm Mg},{\rm Mn},{\rm Co},{\rm Ni}$)~\cite{Kajita2024ChemMater,Kajita2025JPSJ} & $A_z^{\mathrm{loc}}$ & $P\bar3$ & $P\bar3m1$ & \cmark & insulator \\
 & Na$_2$Hf(BO$_3$)$_2$~\cite{Nagai2023Na2HfBO3} & $A_z^{\mathrm{loc}}$ & $P\bar3$ & $P\bar3m1$ & \cmark & insulator \\
 & NASICON-type compounds~\cite{Nagai2023NASICON} & $A_z^{\mathrm{loc}}$ & $P\bar3$ & $P\bar3m1$ & \cmark & insulator \\
 & RbFe(MoO$_4$)$_2$~\cite{Jin_RbFeMoO42_2020,Owen2021PRB,Hayashida2021PRM,Zeng2025Science} & $A_z^{\mathrm{loc}}$ & $P\bar3$ & $P\bar3m1$ & \cmark & insulator \\
 & 1T-TiSe$_2$~\cite{Fu2016TiSe2Strain,Jiang2026TiSe2} & $A_z^{\mathrm{loc}}$ & $P\bar3c1$ & $P\bar3m1$ &  & CDW semimetal \\
 & NiTiO$_3$~\cite{Hayashida2020NatCommun,Hayashida2021PRM,Yokota2022npj,Guo2023PRB,Kusuno2026PRL} & $A_z^{\mathrm{loc}}$ & $R\bar3$ & $R\bar3c$ & \cmark & insulator \\
\addlinespace[0.25em]
Hexagonal & Ba$_3$NaIr$_2$O$_9$~\cite{Kim2009Ba3NaIr2O9,Jain2013MaterialsProject} & $A_z^{\mathrm{loc}}$ & $C2/c$ & $P6_3/mmc$ & \cmark & insulator/correlated oxide \\
 & Ca$_5$Ir$_3$O$_{12}$~\cite{Hasegawa_ferro_rotation_2020,Hanate2021Ca5Ir3O12Superlattice,Hanate2023Ca5Ir3O12SpaceGroup,Hayami2023Ca5Ir3O12Cluster} & $A_z^{\mathrm{loc}}$ & $R3$ & $P\bar6 2m$ &  & semiconductor \\
\addlinespace[0.25em]
Cubic I & CaMn$_7$O$_{12}$~\cite{Bochu1980CaMn7O12,PhysRevLett.108.067201,Yuan2015CaMn7O12Domains,Souliou2016CaMn7O12SoftPhonon} & $A_{[111]}$ & $R\bar3$ & $Im\bar3$ & \cmark & insulator/multiferroic \\
& CdMn$_7$O$_{12}$~\cite{Zhou2023CdMn7O12} & $A_{[111]}$ & $R\bar3$ & $Im\bar3$ & \cmark & insulator/multiferroic \\
\addlinespace[0.25em]
Cubic II & Ba$_2$CaReO$_6$~\cite{YAMAMURA2006Ba2CaReO6,Ishikawa2021Ba2CaReO6,Jain2013MaterialsProject} & $A_{[001]}$ & $I4/m$ & $Fm\bar3m$ & \cmark & correlated $5d^1$ oxide \\
& Ba$_2$HoTaO$_6$~\cite{Kennedy2007Ba2HoTaO6,Jain2013MaterialsProject} & $A_{[001]}$ & $I4/m$ & $Fm\bar3m$ & \cmark & insulator \\
& CaSnF$_6$~\cite{Mayer1983CaSnF6,DayRoberts2025PRL,Jain2013MaterialsProject} & $A_{[111]}$ & $R\bar3$ & $Fm\bar3m$ & \cmark & insulator \\
\bottomrule
\end{tabular}%
}
\begin{minipage}{0.98\textwidth}
\vspace{0.35em}
\scriptsize
$^{\dagger}$ WO$_3$ is listed in the orthorhombic row because the relevant monoclinic phases are reached through an orthorhombic member such as $Pbcn$ in its multi-step phase sequence; tetragonal phases such as $P4/ncc$ also occur at higher temperature and should not be read as a single direct tetragonal-to-monoclinic transition.\\
$^{\ddagger}$ For $R$Te$_3$, $C_{2h}$ denotes the point-group symmetry of the finite-$q$ ferroaxial CDW state, not an ordinary crystallographic space group.
\end{minipage}
\end{table*}

Table~\ref{tab:classification} summarizes the lowest-order static homogeneous strain polynomials that transform as selected ferroaxial components in each non-pyroaxial point-group family.  
These polynomials define the symmetry-allowed strain-derived axial field $h(\bm{\varepsilon})$ entering the conjugate coupling $-h(\bm{\varepsilon})A$.  
For each parent symmetry, we list both the leading basal-plane strain field and a representative leading field involving non-basal strain components.  
The detailed derivation is given in the Supplemental Material~\cite{SupplementalMaterial}.

For the non-cubic crystal families, the local coordinate system is chosen so that the symmetry-distinguished ferroaxial axis coincides with the local $z$ axis, and the corresponding order parameter is denoted by $A_z^{\mathrm{loc}}$. 
This notation is purely local: depending on the crystallographic setting, the same ferroaxial direction may correspond to the conventional $a$, $b$, or $c$ axis.
For the cubic families, the selected ferroaxial component is specified explicitly as $A_{[001]}$, $A_{[110]}$, or $A_{[111]}$ in the parent cubic frame.  
The strain variables carrying the subscript ``loc'' are then defined in the corresponding rotated coordinate system.

Throughout Table~\ref{tab:classification}, the strain variables are expressed in the selected local frame as
\begin{equation}
 X = \varepsilon_{xx} - \varepsilon_{yy}, \qquad Y=2\varepsilon_{xy},
 \label{eq:strain_components}
\end{equation}
with the non-basal components
\begin{equation}
 Z = 2\varepsilon_{zz}-\varepsilon_{xx}-\varepsilon_{yy},\qquad U=2\varepsilon_{xz}, \qquad V=2\varepsilon_{yz}.
 \label{eq:nonbasal_components}
\end{equation}
For a pure in-plane uniaxial deviatoric strain with signed amplitude $\varepsilon_{\rm u}$ and principal-axis angle $\theta$,
\begin{equation}
 X = 2\varepsilon_{\rm u}\cos2\theta,
 \qquad
 Y = 2\varepsilon_{\rm u}\sin2\theta,
 \label{eq:uniaxial_XY}
\end{equation}
which directly yields the angular dependence listed in the leading-basal-field column of Table~\ref{tab:classification}.

The resulting hierarchy follows a remarkably systematic pattern.
Orthorhombic parents allow the linear shear field $Y$.  
Tetragonal parents require the quadratic field $XY$.  
Trigonal and hexagonal parents require the cubic basal field $Y(3X^2-Y^2)$, giving the characteristic $\sin6\theta$ dependence under in-plane uniaxial deviatoric strain.  
Cubic parents require an additional axis specification, and different projections, such as $[001]$, $[110]$, and $[111]$, therefore lead to different local strain channels.

The subgroup columns indicate the residual point group of a representative strained tensor that activates the corresponding axial field.  
The residual symmetry is determined by the complete strain tensor rather than by the scalar strain polynomial alone.
For example, although the polynomial $XY$ remains invariant under an operation that maps $(X,Y)$ to $(-X,-Y)$, the strain tensor itself changes under this operation, so it is not retained as a symmetry of the strained crystal.

Representative material platforms corresponding to the symmetry classes in Table~\ref{tab:classification} are summarized in Table~\ref{tab:materials}.
The listed systems include both proper ferroaxial transitions, in which the ferroaxial order itself is the primary $\Gamma$-point order parameter, and improper cases where a uniform ferroaxial moment emerges as a secondary composite order induced by a finite-wave-vector structural or electronic instability, such as CDW or superlattice formation.
In the latter case, the strain-derived field couples to the induced uniform ferroaxial component rather than the primary finite-$\bm{Q}$ order parameter as given by Eq.~(\ref{eq:finite_q_composite}).
Table~\ref{tab:materials} therefore provides a symmetry-based guide to representative ferroaxial material platforms, rather than a complete list of proper $\Gamma$-point ferroaxial transitions.

%%%%%%%%%%%%%%%%%%%%%%%%%%%%%%%%%%%%%%%%%%%%%%%%%%
\section{First-principles demonstration in trigonal Na$_2$BaMg(PO$_4$)$_2$}
\label{sec:demo}

In this section, we demonstrate the piezoaxial coupling by first-principles calculations for the trigonal ferroaxial compound Na$_2$BaMg(PO$_4$)$_2$.
This glaserite-type material provides a prototypical structural ferroaxial system in which the order parameter $A$ is the collective rotation of PO$_4$ tetrahedra about the trigonal axis.
We verify the symmetry-predicted basal-plane field $h(\bm{\varepsilon})\propto Y(3X^2-Y^2)$ through three complementary calculations: clamped-coordinate domain splittings, biased double-well scans along the ferroaxial rotation coordinate, and fixed-cell relaxations from the para-axial structure.  
These calculations confirm the characteristic $\varepsilon_{\rm u}^3\sin6\theta$ dependence, its sign reversal under strain-axis rotation or strain-sign reversal, and the resulting strain-selected ferroaxial domain preference.

%%%%%%%%%%
\subsection{Trigonal glaserite case and strain convention}
\label{sec:trigonal}

Na$_2$BaMg(PO$_4$)$_2$ is a useful minimal target because its ferroaxial transition is a structural displacive-type transition in the glaserite family, whose ferroaxial order parameter can be represented by a rotation angle $\phi$ of PO$_4$ tetrahedra about the trigonal axis~\cite{Yamagishi2023Glaserite,Kajita2024ChemMater,Kajita2025JPSJ}, 
\begin{equation}
 \phi = \tan^{-1} \left( \frac{\sum_{i = 1}^3(\bm{r}_{i}^{0} \times \bm{r}_{i})_{z}}{\sum_{i=1}^{3} \bm{r}_{i}^{0} \cdot \bm{r}_{i}} \right).
 \label{eq:phi_def}
\end{equation}
Here, $\bm{r}_{i}^{0}$ is the in-plane P--O bond vector of the $i$-th basal oxygen in the para-axial reference structure and $\bm{r}_{i}$ is the corresponding vector in the distorted structure.
The two PO$_4$ tetrahedra in the primitive cell have the same ferroaxial rotation sign as shown in Fig.~\ref{fig:concept}, so the two ferroaxial domains are distinguished simply by the sign of $\phi$.

The high-temperature para-axial phase with $\phi = 0$ belongs to space group $P\bar{3}m1$ with point group $\bar{3}m$ ($D_{3d}$), whereas the ferroaxial phase with $\phi \sim 9.06^{\circ}$ at around room temperature below the transition temperature $T_{\rm c}\simeq540$ K belongs to space group $P\bar{3}$ with point group $\bar{3}$ ($C_{\rm 3i}$)~\cite{Kajita2024ChemMater}.
The single ferroaxial component considered here is the $c$-axis component $A \propto \phi$, which transforms as the $A_{2g}$ irreducible representation of the parent point group $\bar{3}m$ ($D_{3d}$).

The basal-plane strain doublet $(X,Y)$ transforms as the $E_g$ representation.
As discussed in Sec.~\ref{sec:point_group_hierarchy}, neither the first nor the second symmetric power of this doublet contains the $A_{2g}$ axial channel.
The leading symmetry-allowed basal-plane field is therefore cubic,
\begin{equation}
 h(\bm{\varepsilon}) \propto Y(3X^2-Y^2).
 \label{eq:cubic_field}
\end{equation}
This equation is the trigonal specialization of the general field $h(\bm{\varepsilon})$ in Eq.~(\ref{eq:h_expansion}).

For the calculations below, we impose a traceless in-plane deviatoric strain whose two principal axes are rotated by an angle $\theta$ and $\theta + 90^{\circ}$ from the crystallographic $x$-axis, as shown in Fig.~\ref{fig:concept}.
The two in-plane principal strains are $+\varepsilon_{\rm u}$ and $-\varepsilon_{\rm u}$, giving
\begin{equation}
 X = 2 \varepsilon_{\rm u} \cos2\theta, \qquad Y = 2 \varepsilon_{\rm u} \sin2\theta.
 \label{eq:uniaxial_xy}
\end{equation}
This is the same signed-principal-strain convention as Eq.~(\ref{eq:uniaxial_XY}).
Substituting Eq.~(\ref{eq:uniaxial_xy}) into Eq.~(\ref{eq:cubic_field}), we obtain
\begin{equation}
 h(\varepsilon_{\rm u},\theta) \propto \varepsilon_{\rm u}^3\sin6\theta.
 \label{eq:sin6}
\end{equation}
Thus the field changes sign either when the tensile and compressive principal strains are interchanged, $\varepsilon_{\rm u}\rightarrow-\varepsilon_{\rm u}$, or when the principal axis is rotated by $30^\circ$.
Accordingly, $\theta=0^\circ$, $30^\circ$, $60^\circ$, and symmetry-equivalent directions are nodal directions where the basal-plane field vanishes.

%%%%%%%%%%
\subsection{Computational methods}
\label{sec:methods}

First-principles calculations were performed using \textsc{Quantum ESPRESSO}~\cite{Giannozzi2009QE,Giannozzi2017QE}.
We used the PBEsol exchange-correlation functional~\cite{Perdew2008PBEsol} and scalar-relativistic optimized norm-conserving Vanderbilt pseudopotentials~\cite{ONCV_PRB_2013} downloaded from PseudoDojo~\cite{PseudoDojo_2018}.
The kinetic energy cutoff of the Kohn-Sham orbitals and the convergence threshold are set to be 100 Ry and $1 \times 10^{-14}$ Ry, and the $\bm{k}$ grid is taken as $(N_{1}, N_{2}, N_{3}) = (8, 8, 6)$.

Starting from the relaxed unstrained structure, we impose the traceless in-plane deviatoric strain tensor
\begin{equation}
 \bm{\varepsilon}(\varepsilon_{\rm u},\theta)=\varepsilon_{\rm u}
 \begin{pmatrix}
 \cos 2\theta & \sin 2\theta & 0\\
 \sin 2\theta & -\cos 2\theta & 0\\
 0 & 0 & 0
 \end{pmatrix},
 \label{eq:strain_matrix_methods}
\end{equation}
or equivalently,
\begin{equation}
 \bm{\varepsilon} = \varepsilon_{\rm u} (\hat{\bm n}\otimes\hat{\bm n} - \hat{\bm m}\otimes\hat{\bm m}),
\end{equation}
where $\hat{\bm n} = (\cos\theta,\sin\theta,0)$ and $\hat{\bm m} = (-\sin\theta,\cos\theta,0)$.
For $\varepsilon_{\rm u} > 0$, the crystal is stretched along $\hat{\bm n}$ and compressed along $\hat{\bm m}$; for $\varepsilon_{\rm u} < 0$, these two directions are interchanged.
A literal one-axis deformation contains, in addition, an isotropic basal-plane component, which changes the elastic background but does not affect the domain-odd angular dependence.
With the definitions in Eqs.~(\ref{eq:strain_components}) and (\ref{eq:nonbasal_components}), Eq.~(\ref{eq:strain_matrix_methods}) gives exactly Eq.~(\ref{eq:uniaxial_XY}).

The strained cell $\mathbf{L}'$ and initially strained Cartesian coordinates $\bm{r}'_{i}$ are generated by the affine transformation
\begin{equation}
 \mathbf{L}' = \mathbf{L}(\mathbf{I} + \bm{\varepsilon})^{\rm T},
 \qquad
 \bm{r}'_{i} = \bm{r}_{i}(\mathbf{I} + \bm{\varepsilon})^{\rm T}.
 \label{eq:affine_strain}
\end{equation}
Here, the rows of $\mathbf{L}$ are the non-strained lattice vectors, and $\bm{r}_{i}$ is the Cartesian row vector of the $i$th atom without strain.
For internal-coordinate relaxations, the strained cell was then fixed and only atomic positions were relaxed.

To verify the piezoaxial coupling, we perform three complementary calculations.
First, clamped-coordinate domain splittings are evaluated by applying the same strained cell to the two preconstructed ferroaxial states with opposite sign of $\phi$ and computing
\begin{equation}
 \Delta E_{\rm dom} = E(+|\phi|) - E(-|\phi|).
 \label{eq:domain_split_dft}
\end{equation}
Second, one-dimensional scans along the ferroaxial coordinate $\phi$ are used to show directly that strain tilts the double-well potential.
Third, fixed-cell relaxations are started from the para-axial $\phi = 0$ structure at finite strain, testing whether the strain field selects a ferroaxial basin without imposing an initial nonzero rotation.

%%%%%%%%%%
\subsection{Clamped-coordinate verification of the cubic field}
\label{sec:clamped}

\begin{figure*}[t]
\includegraphics[width=0.98\textwidth]{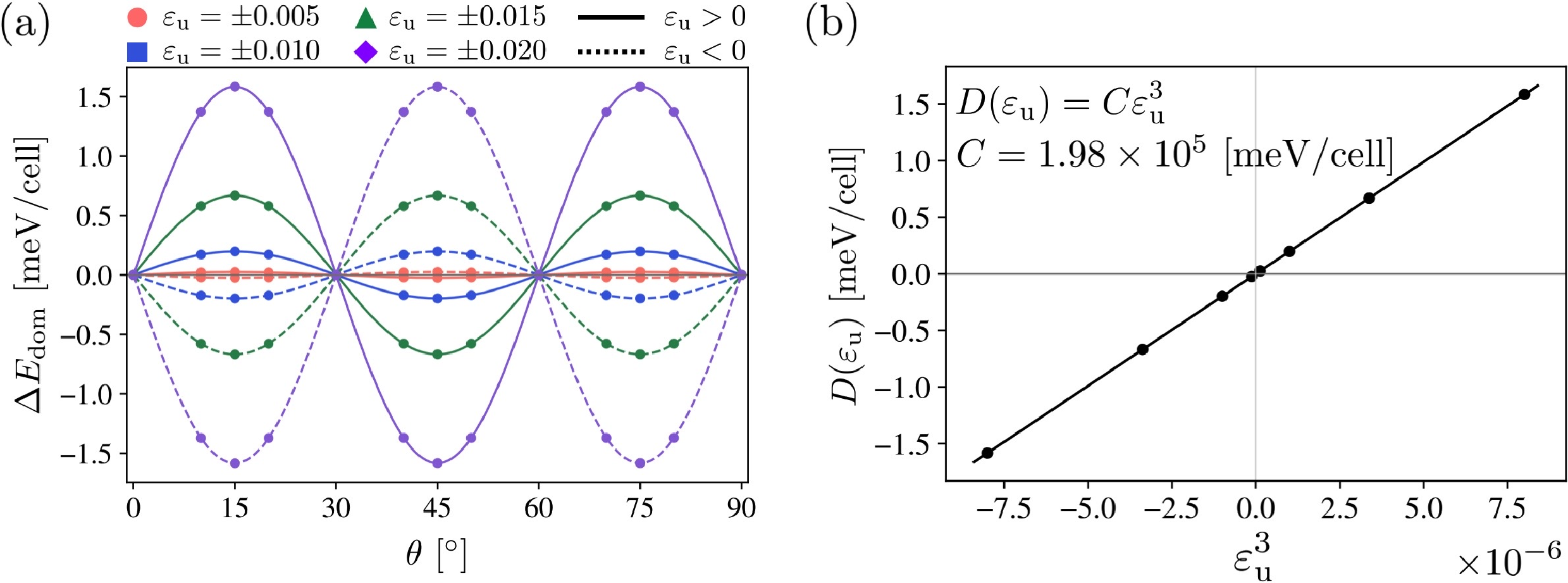}
\caption{
Clamped-coordinate verification of the signed cubic piezoaxial field in Na$_2$BaMg(PO$_4$)$_2$.
(a) Angular dependence of the domain splitting $\Delta E_{\rm dom} = E(+|\phi|) - E(-|\phi|)$ for signed deviatoric strains $\varepsilon_{\rm u} = \pm 0.005$, $\pm 0.010$, $\pm 0.015$, and $\pm 0.020$.
The points are first-principles total-energy differences, and the curves are fits constrained to the $\sin6\theta$ form; solid and dashed lines denote $\varepsilon_{\rm u} > 0$ and $\varepsilon_{\rm u}<0$, respectively.
(b) Fitted signed amplitudes $D(\varepsilon_{\rm u})$ as a function of $\varepsilon_{\rm u}^{3}$.
The black line is a fit constrained through the origin, $D(\varepsilon_{\rm u}) = C\varepsilon_{\rm u}^{3}$, with $C = 1.9792\times10^5$ meV/cell.
}
\label{fig:scaling}
\end{figure*}

For each signed strain amplitude and strain angle, we compute the clamped-coordinate domain splitting in Eq.~(\ref{eq:domain_split_dft}).
The symmetry prediction is
\begin{equation}
 \Delta E_{\rm dom}^{\rm fit}(\varepsilon_{\rm u},\theta) = D(\varepsilon_{\rm u})\sin6\theta,
 \label{eq:scf_law}
\end{equation}
with a signed amplitude satisfying $D(-\varepsilon_{\rm u}) = -D(+\varepsilon_{\rm u})$.
Figure~\ref{fig:scaling}(a) shows that the calculated splittings follow this angular dependence for all four strain magnitudes and for both strain signs.
The sign reversal between positive and negative $\varepsilon_{\rm u}$ is particularly important: it rules out an even-in-strain elastic origin for the domain-odd splitting.

Figure~\ref{fig:scaling}(b) confirms the strain-amplitude dependence.
A fit constrained through the origin gives
\begin{equation}
D(\varepsilon_{\rm u}) = C\varepsilon_{\rm u}^3, \qquad C=1.9792\times10^5~{\rm meV}/{\rm cell}.
\label{eq:cubic_fit}
\end{equation}
The angular fits and the signed cubic scaling verify that the calculated domain splitting is governed by the trigonal cubic invariant in Eqs.~(\ref{eq:cubic_field}) and (\ref{eq:sin6}).

%%%%%%%%%%
\subsection{Bias of the ferroaxial double well}
\label{sec:double_well}
\label{sec:phiscans}

\begin{figure*}[t]
\includegraphics[width=0.98\linewidth]{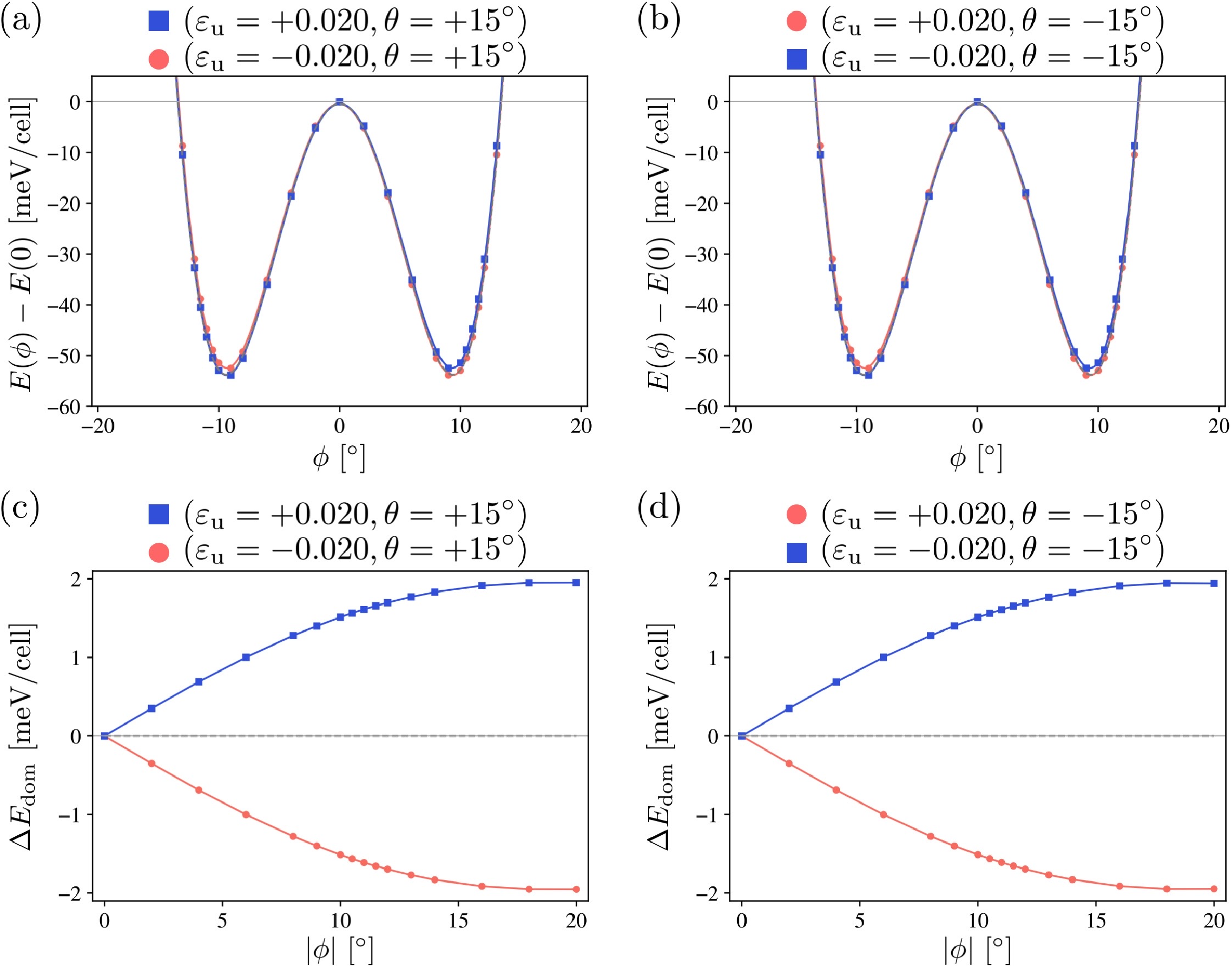}
\caption{
Biased ferroaxial double-well potentials and the corresponding domain splittings obtained from PO$_4$ rotation scans.
(a),(b) Relative energy $E(\phi)-E(0)$ as a function of the signed rotation angle $\phi$ for $\theta=+15^\circ$ and $\theta=-15^\circ$, respectively.
Points are first-principles energies and solid curves are sixth-order polynomial fits used as guides to the eye; the gray dashed curve is the zero-strain reference.
(c),(d) Corresponding domain splitting $E(+|\phi|)-E(-|\phi|)$ for $\theta=+15^\circ$ and $\theta=-15^\circ$, respectively.
The sign reversal upon changing either $\varepsilon_{\rm u}$ or $\theta$ is the double-well manifestation of the signed cubic field $h\propto\varepsilon_{\rm u}^3\sin6\theta$.
}
\label{fig:phibias}
\end{figure*}

The clamped-coordinate comparison verifies the symmetry of the domain splitting.
We next confirm that the same strain-derived field appears as an actual tilt of the ferroaxial double-well potential.
Figure~\ref{fig:phibias} shows the energy profile as a function of the PO$_4$ rotation angle $\phi$ for $|\varepsilon_{\rm u}|=0.020$ and $\theta=\pm15^\circ$, together with the corresponding odd-in-$\phi$ domain splitting.
Without strain, the double well is symmetric, as required by the degeneracy of the two mirror-related ferroaxial domains.
Applying strain introduces an odd-in-$\phi$ contribution that biases one minimum over the other.
Reversing either the strain sign ($\varepsilon_{\rm u}$) or the strain orientation ($\theta=\pm15^\circ$) reverses this odd contribution, exactly as expected from Eq.~(\ref{eq:sin6}).

This behavior clearly distinguishes the piezoaxial coupling from conventional strain couplings that are even in the ferroaxial order parameter.
Terms such as $A^2\varepsilon_{\mu}$ can shift the curvature, transition temperature, or barrier height, but they do not lift the degeneracy between $+|\phi|$ and $-|\phi|$.
The odd component of the profile in Fig.~\ref{fig:phibias} therefore provides a direct microscopic realization of the ferroaxial coupling $-h(\bm{\varepsilon})A$.

%%%%%%%%%%
\subsection{Fixed-strain relaxation from the para-axial structure}
\label{sec:relax_summary}

\begin{figure}[t]
\includegraphics[width=0.48\textwidth]{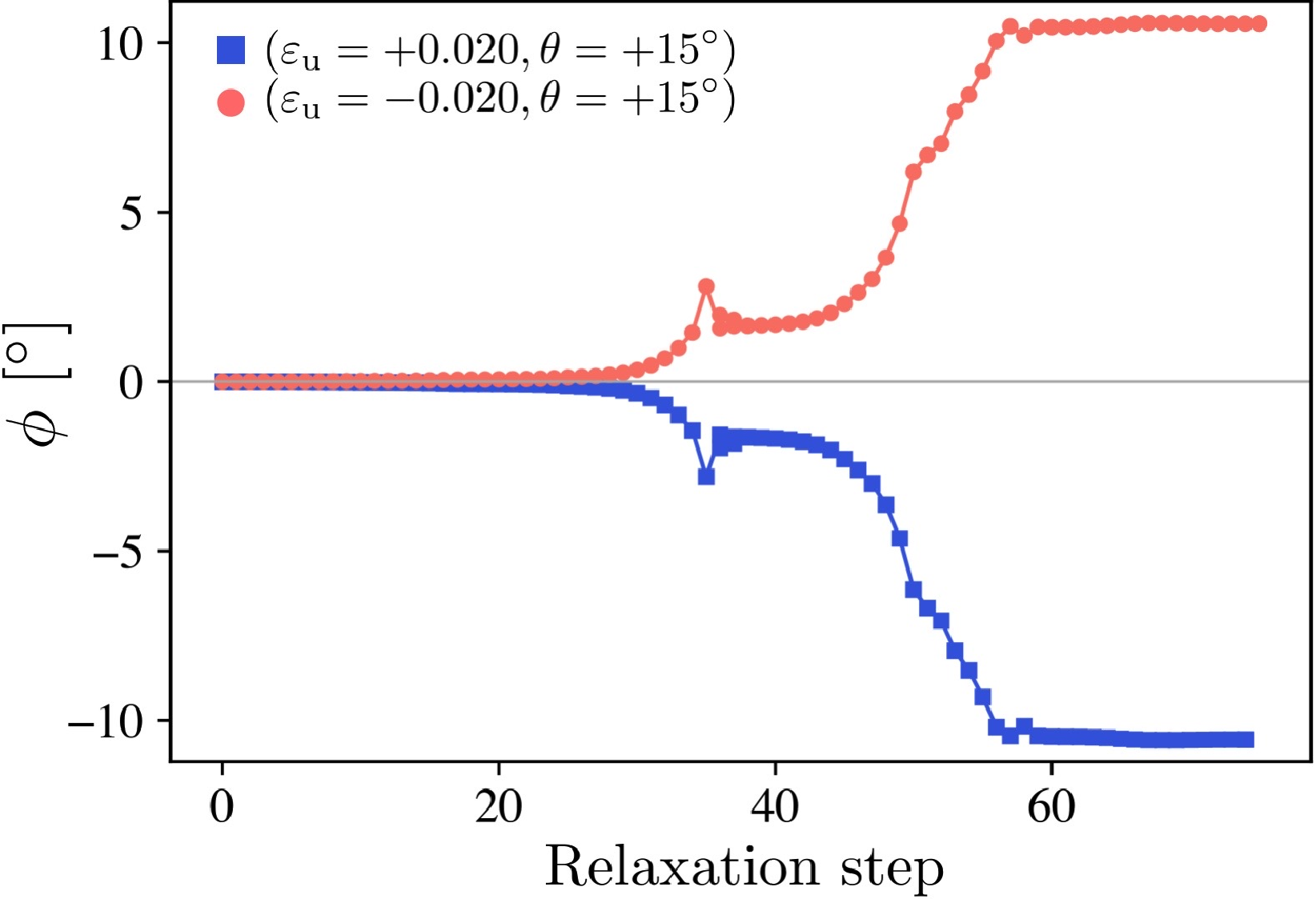}
\caption{
Fixed-strain relaxation from the para-axial structure.
The plotted quantity is the average PO$_4$ rotation angle $\phi$, given by Eq.~(\ref{eq:phi_def}), extracted from each ionic relaxation step for $\varepsilon_{\rm u} = +0.020$ (blue) and $\varepsilon_{\rm u} = -0.020$ (red) at $\theta = +15^\circ$.
Starting from the same para-axial configuration with $\phi = 0$, the two strain signs relax into opposite ferroaxial basins, demonstrating strain-selective domain formation.
}
\label{fig:relax}
\end{figure}

Next, we show the results of fixed-cell relaxations from the para-axial structure with $\phi = 0$ after applying the strained cell.
Unlike the previous calculations, no initial ferroaxial distortion is introduced, making this a more stringent test of whether the strain field alone selects the ferroaxial domain.
We use $\theta = +15^\circ$, for which $\sin6\theta = 1$, so the sign of the domain bias is reversed by reversing $\varepsilon_{\rm u}$.

Figure~\ref{fig:relax} shows the relaxation trajectory at $|\varepsilon_{\rm u}| = 0.020$.
Although both calculations start from the same para-axial configuration, they relax into opposite ferroaxial minima: $\varepsilon_{\rm u} = +0.020$ relaxes toward $\phi < 0$, whereas $\varepsilon_{\rm u} = -0.020$ relaxes toward $\phi > 0$.
The strain field therefore determines the ferroaxial basin reached during structural relaxation, providing a direct real-space demonstration of strain-selective ferroaxial domain formation.

%%%%%%%%%%
\subsection{Estimate for strain-cooling experiments}
\label{sec:straincooling}

\begin{table}[t]
\caption{
Clamped-coordinate signed amplitudes in Eq.~(\ref{eq:scf_law}) with $\theta = 15^\circ$
The positive amplitude $D_0(|\varepsilon_{\rm u}|)$ is defined by $D(\varepsilon_{\rm u}) = {\rm sgn}(\varepsilon_{\rm u})D_0(|\varepsilon_{\rm u}|)$.
}
\label{tab:scfamps}
\centering
\renewcommand{\arraystretch}{1.18}
\begin{ruledtabular}
\begin{tabular}{ccc}
$|\varepsilon_{\rm u}|$ & $D_0(|\varepsilon_{\rm u}|)$ (meV/cell) & $D_0/k_{\rm B}$ (K/cell) \\
\hline
0.005 & 0.0247 & 0.287\\
0.010 & 0.198 & 2.30 \\
0.015 & 0.668 & 7.75 \\
0.020 & 1.58 & 18.4 \\
\end{tabular}
\end{ruledtabular}
\end{table}

Finally, we estimate the magnitude of the strain bias relevant to strain-field-cooling experiments.
The situation is analogous to conventional magnetic field cooling, where the Zeeman energy $- B M$ biases magnetic-domain formation.
Here, the corresponding free-energy bias is
\begin{equation}
 F_{\rm A}^{\rm bias} = -h(\bm{\varepsilon}) A,
 \label{eq:fa_field_cooling}
\end{equation}
with $h(\bm{\varepsilon}) \propto \varepsilon_{\rm u}^3\sin6\theta$ for the trigonal basal-plane strain considered above.

The experimental protocol suggested by Fig.~\ref{fig:concept} is therefore straightforward: apply a fixed in-plane deviatoric strain above $T_{\rm c}$, cool through the $P\bar{3}m1 \rightarrow P\bar{3}$ transition, and image the resulting ferroaxial domains, for example by linear electrogyration~\cite{Yamagishi2023Glaserite,Kajita2025JPSJ}.
The most important internal controls are symmetry controls: a $30^\circ$ rotation of the strain axis should reverse the selected domain, a $60^\circ$ rotation should restore the same selection, and a null direction such as $\theta = 0^\circ$ should suppress the domain bias associated with the cubic basal-plane field.

From Table~\ref{tab:scfamps}, the maximal one-cell splittings are
\begin{equation}
 \frac{D_0(0.010)}{k_{\rm B}} = 2.30~{\rm K/cell},
 \qquad
 \frac{D_0(0.020)}{k_{\rm B}} = 18.4~{\rm K/cell}.
 \label{eq:d_examples}
\end{equation}
Although these one-cell energy splittings are much smaller than $T_{\rm c}$, ferroaxial domains nucleate collectively rather than cell by cell.
For a correlated nucleus containing $N$ primitive cells, a simple population estimate gives
\begin{equation}
\frac{P_{\rm fav}}{P_{\rm unfav}} \simeq \exp\left[\frac{ND_0(|\varepsilon_{\rm u}|)}{k_{\rm B}T}\right].
 \label{eq:boltzmann_ratio}
\end{equation}
Using $T = T_{\rm c}\simeq540$ K~\cite{Kajita2024ChemMater}, the exponent is about 3.4 for $|\varepsilon_{\rm u}| = 0.020$ and $N = 100$, and about 2.1 for $|\varepsilon_{\rm u}| = 0.010$ and $N = 500$.
Although this estimate does not replace a finite-temperature nucleation theory, it suggests that the strain-induced bias can become thermodynamically significant when accumulated over a mesoscopic correlated volume.

%%%%%%%%%%%%%%%%%%%%%%%%%%%%%%%%%%%%%%%%%%%%%%%%%%
\section{Scope and related material classes}
\label{sec:scope}

The first-principles calculations of this work is deliberately restricted to the prototypical trigonal ferroaxial compound Na$_2$BaMg(PO$_4$)$_2$, where the ferroaxial order parameter is a simple structural rotation of PO$_4$ tetrahedra.
For such a system, the calculated domain-odd energy splitting can be interpreted directly as the action of a homogeneous strain-derived conjugate field.
By contrast, the broader symmetry classification in Table~\ref{tab:classification} and the representative materials listed in Table~\ref{tab:materials} should be regarded as a general symmetry framework rather than a list of systems expected to exhibit equally direct first-principles verification.

The most straightforward targets are proper structural ferroaxial transitions.
Trigonal glaserite-type phosphates and related $P\bar{3}m1\rightarrow P\bar{3}$ systems are especially suitable because the ferroaxial order parameter is a structural rotation about a unique trigonal axis and the leading basal-plane field exhibits the characteristic $\sin6\theta$ dependence.
Centrosymmetric cubic-to-rhombohedral examples such as CaMn$_7$O$_{12}$, CdMn$_7$O$_{12}$, and CaSnF$_6$ are also useful symmetry targets, although the relevant local axial direction and the allowed strain order must be chosen according to the cubic classifications of Table~\ref{tab:classification}.
Orthorhombic and tetragonal systems illustrate lower-order strain fields, including linear and quadratic basal-plane couplings.
Their microscopic physics, however, is often complicated by electronic correlations, metal--insulator transitions, heavy-fermion behavior, or coupled lattice-electronic instabilities, as exemplified by Ta$_2$NiSe$_5$, $R$NiO$_3$, CeCu$_6$, VO$_2$, and several iridates.
Although the same symmetry-allowed invariant exists in these systems, quantitatively extracting a piezoaxial coefficient generally requires disentangling the strain-induced axial field from additional material-specific electronic and structural effects.

Finite-wave-vector and electronic examples require a different distinction.
Since translational symmetry forbids a homogeneous strain from coupling linearly to a primary finite-$\bm{Q}$ order parameter, the strain field instead biases a uniform axial composite induced by the finite-$\bm{Q}$ order, as discussed in Eq.~(\ref{eq:finite_q_composite}).
This interpretation applies naturally to 1T-TiSe$_2$, rare-earth tritellurides, and Ca$_5$Ir$_3$O$_{12}$.
In 1T-TiSe$_2$, the ferroaxial character is associated with the CDW state; in $R$Te$_3$, the $C_{2h}$ entry in Table~\ref{tab:materials} denotes the point-group symmetry of the electronic ferroaxial density-wave state rather than an ordinary three-dimensional crystallographic space group~\cite{Singh2025RTe3}; and in Ca$_5$Ir$_3$O$_{12}$, the electric-toroidal order appears with a $\bm{q} = (1/3,1/3,1/3)$ superlattice and a low-temperature $R3$ structure~\cite{Hanate2021Ca5Ir3O12Superlattice,Hanate2023Ca5Ir3O12SpaceGroup,Hayami2023Ca5Ir3O12Cluster}.
These examples demonstrate that the present symmetry framework extends beyond simple structural ferroaxial transitions, provided that the strain field is understood as coupling to the induced uniform axial degree of freedom.

A further caveat concerns symmetry lowering that is not purely ferroaxial.  
The piezoaxial coupling requires only that the strain polynomial and the targeted axial component belong to the same irreducible representation of the parent point group.  
Consequently, polar, chiral, or other symmetry-breaking order parameters may coexist with ferroaxiality without invalidating the coupling itself.
Such systems remain valuable platforms for exploring strain-controlled ferroaxial phenomena, although additional symmetry breaking may obscure the interpretation of the measured response.
For this reason, centrosymmetric proper ferroaxial transitions continue to provide the clearest benchmark for experimentally establishing homogeneous strain as a static conjugate field for ferroaxial order.

%%%%%%%%%%%%%%%%%%%%%%%%%%%%%%%%%%%%%%%%%%%%%%%%%%
\section{Conclusions}
\label{sec:conclusion}

We have formulated a symmetry-based piezoaxial coupling in which static homogeneous strain generates an axial field conjugate to a selected ferroaxial component.  
In the single-component notation, the coupling is written as $-h(\bm{\varepsilon})A$, where the allowed strain polynomial is uniquely determined by the parent point-group symmetry and the chosen ferroaxial axis.
For basal-plane strain, the leading field is linear in orthorhombic parents, quadratic in tetragonal parents, and cubic in trigonal and hexagonal parents.  
Cubic parents require an explicit axis specification because different projections, such as $[001]$, $[110]$, and $[111]$, lead to different local strain channels.

We verified the trigonal case by first-principles calculations for Na$_2$BaMg(PO$_4$)$_2$, a glaserite-type ferroaxial material whose order parameter is represented by the rotation of PO$_4$ tetrahedra about the trigonal axis.  
The clamped-coordinate domain splitting follows the symmetry-predicted angular dependence $\Delta E_{\rm dom} = D(\varepsilon_{\rm u})\sin6\theta$, with $D(\varepsilon_{\rm u})\propto\varepsilon_{\rm u}^3$.  
Furthermore, PO$_4$ rotation scans show that the strain-derived field tilts the ferroaxial double-well potential, and fixed-cell relaxations from the para-axial structure evolve into opposite ferroaxial minima depending solely on the sign of the applied strain.

These results establish static homogeneous strain as a symmetry-allowed conjugate field for ferroaxial domain selection and provide a concrete strain-field-cooling protocol based on symmetry.
Reversing the sign of the deviatoric strain or rotating the in-plane strain axis by $30^\circ$ reverses the preferred ferroaxial domain, while a $60^\circ$ rotation restores the same domain selection.  
Although the most direct realization is provided by proper structural ferroaxial transitions, the same symmetry framework naturally extends to electronic or finite-wave-vector axial ferroaxial states by interpreting the strain field as coupling to an induced uniform axial composite.
The present work therefore establishes a general symmetry framework for strain-controlled ferroaxiality and provides a microscopic foundation for ferroaxial domain engineering and ferroaxiality-driven functionalities.

%%%%%%%%%%%%%%%%%%%%%%%%%%%%%%%%%%%%%%%%%%%%%%%%%%
\begin{acknowledgments}
This work was supported by JSPS KAKENHI Grants Numbers JP22H00101, JP22H01183, JP23H04869, JP23K03288, JP26H00618, JP26K17075, and by JST CREST (JPMJCR23O4) and JST FOREST (JPMJFR2366).
\end{acknowledgments}

%%%%%%%%%%%%%%%%%%%%%%%%%%%%%%%%%%%%%%%%%%%%%%%%%%
\bibliographystyle{apsrev4-2}
\bibliography{ref}

%%%%%%%%%%%%%%%%%%%%%%%%%%%%%%%%%%%%%%%%%%%%%%%%%%
\clearpage
\onecolumngrid

% ===== Supplemental Material =====

\setcounter{section}{0}
\setcounter{subsection}{0}
\setcounter{equation}{0}
\setcounter{figure}{0}
\setcounter{table}{0}

\renewcommand{\thesection}{S\arabic{section}}
\renewcommand{\thesubsection}{S\arabic{section}.\arabic{subsection}}
\renewcommand{\theequation}{S\arabic{equation}}
\renewcommand{\thefigure}{S\arabic{figure}}
\renewcommand{\thetable}{S\arabic{table}}

\begin{center}
{\large \bf Supplemental Material for\\
``Piezoaxial coupling for strain-selected ferroaxial domain control''}\\[1em]
Rikuto Oiwa and Satoru Hayami
\end{center}

\section{Symmetry setting and notation}

We derive homogeneous strain fields that can linearly bias a selected ferroaxial component.
The selected ferroaxial direction is denoted by the local $z$ axis, and the coupling is written as
\begin{equation}
 F_A=-h_z^{\mathrm{loc}}(\bm\varepsilon)A_z^{\mathrm{loc}}.
 \label{eq:coupling}
\end{equation}
Here $A_z^{\mathrm{loc}}$ is the selected component of the axial, or pseudovector, ferroaxial order parameter, and $h_z^{\mathrm{loc}}(\bm\varepsilon)$ is the conjugate field generated by a homogeneous strain tensor $\bm\varepsilon$.
The superscript ``loc'' emphasizes that the local $z$ axis is chosen along the ferroaxial direction; it is not necessarily the conventional crystallographic $c$ axis.

For an orthogonal point-group operation $g$, a polar vector transforms as $\bm v\mapsto g\bm v$, whereas an axial vector transforms as
\begin{equation}
 \bm A\mapsto D^{\rm ax}(g)\bm A,
 \qquad
 D^{\rm ax}(g)=\det(g)g .
 \label{eq:axial_rep}
\end{equation}
The strain tensor is a symmetric polar rank-two tensor and transforms as
\begin{equation}
 \bm\varepsilon\mapsto g\bm\varepsilon g^{\mathsf T}.
 \label{eq:strain_transform}
\end{equation}
Therefore a strain-induced axial-vector field must satisfy the covariance condition
\begin{equation}
 \bm h(g\bm\varepsilon g^{\mathsf T})=D^{\rm ax}(g)\bm h(\bm\varepsilon).
 \label{eq:vector_covariance}
\end{equation}
Equation~\eqref{eq:vector_covariance} is the most general form and is especially important for cubic point groups, where symmetry operations can permute the Cartesian components of $\bm A$.
For non-cubic rows, the selected local $z$ axis is symmetry-distinguished and $A_z^{\mathrm{loc}}$ spans a one-dimensional representation. In that case Eq.~\eqref{eq:vector_covariance} reduces to
\begin{equation}
 h_z^{\mathrm{loc}}(g\bm\varepsilon g^{\mathsf T})=\chi_{A_z^{\mathrm{loc}}}(g)h_z^{\mathrm{loc}}(\bm\varepsilon),
 \qquad
 \chi_{A_z^{\mathrm{loc}}}(g)=\det(g)g_{zz},
 \label{eq:scalar_covariance}
\end{equation}
where $g_{zz}$ is the $zz$ component of $g$ in the local frame.
Equation~\eqref{eq:scalar_covariance} should not be used as a full-cubic-group character of a fixed cubic component, because a cubic operation such as a threefold rotation about $[111]$ can cyclically permute $A_x,A_y,A_z$.
For cubic rows we instead construct a cubic-covariant vector field using Eq.~\eqref{eq:vector_covariance} and then project it onto the selected local axis.

The local strain coordinates used throughout this Supplement are
\begin{align}
T&=\varepsilon_{xx}+\varepsilon_{yy}+\varepsilon_{zz},
&X&=\varepsilon_{xx}-\varepsilon_{yy},
&Y&=2\varepsilon_{xy},
\label{eq:def_TXY}
\\
Z&=2\varepsilon_{zz}-\varepsilon_{xx}-\varepsilon_{yy},
&U&=2\varepsilon_{xz},
&V&=2\varepsilon_{yz}.
\label{eq:def_ZUV}
\end{align}
Here $T=\mathrm{Tr}\varepsilon$ is the trace strain, $(X,Y)$ is the basal-plane quadrupolar strain doublet, $Z$ is the tetragonal normal-strain component with respect to the selected local $z$ axis, and $(U,V)$ are the shears involving the selected local $z$ axis.
Unless a superscript $\mathrm{cub}$ is written explicitly, all strain components are local-frame components.
The inverse relation is
\begin{equation}
\bm\varepsilon(q)=
\begin{pmatrix}
T/3+X/2-Z/6 & Y/2 & U/2\\
Y/2 & T/3-X/2-Z/6 & V/2\\
U/2 & V/2 & T/3+Z/3
\end{pmatrix}.
\label{eq:strain_matrix}
\end{equation}
Here $q=(T,X,Y,Z,U,V)$ denotes the strain-coordinate vector.
The trace strain $T$ is invariant under all point-group operations. Multiplying an allowed field by $T^n$ therefore produces only a higher-order descendant of the same symmetry, and $T$ is omitted when identifying leading fields.

We call a field \emph{basal} if it depends only on $X$ and $Y$.
A field is called \emph{non-basal} if it contains at least one of $Z,U,V$.
The phrase ``leading field'' means the lowest-order homogeneous polynomial in strain that satisfies the appropriate covariance condition.

\section{Derivation of the strain fields}

For non-cubic rows, we use the homogeneous polynomial ansatz
\begin{equation}
 h_z^{(n)}(q)=\sum_\alpha c_\alpha M_\alpha^{(n)}(q),
 \label{eq:poly_ansatz}
\end{equation}
where $M_\alpha^{(n)}(q)$ denotes a monomial of total degree $n$ in the chosen strain variables $q$.
For example, for a basal field $q=(X,Y)$, the degree-one monomials are
\begin{equation}
 M_\alpha^{(1)}=X,\;Y,
\end{equation}
and the degree-two monomials are
\begin{equation}
 M_\alpha^{(2)}=X^2,\;XY,\;Y^2.
\end{equation}
For a non-basal field we use $q=(X,Y,Z,U,V)$ and keep only monomials containing at least one of $Z,U,V$; examples at second order are $XZ$, $YZ$, $U^2$, $UV$, and $V^2$.
Let $\rho_\varepsilon(g)$ denote the linear representation induced by Eq.~\eqref{eq:strain_transform} on the coordinate vector $q$. Explicitly,
\begin{equation}
 g\bm\varepsilon(q)g^{\mathsf T}=\bm\varepsilon(\rho_\varepsilon(g)q).
 \label{eq:rho_eps_def}
\end{equation}
Substituting Eq.~\eqref{eq:poly_ansatz} into Eq.~\eqref{eq:scalar_covariance} gives
\begin{equation}
 \sum_\alpha c_\alpha
 \left[M_\alpha^{(n)}(\rho_\varepsilon(g)q)-\chi_{A_z^{\mathrm{loc}}}(g)M_\alpha^{(n)}(q)\right]=0
 \label{eq:linear_constraint}
\end{equation}
for each generator $g$ of the parent point group.
Equation~\eqref{eq:linear_constraint} is a polynomial identity in the independent strain coordinates, not an equation imposed at one particular strain value.
After expanding it, we collect the result in a monomial basis $N_\beta^{(n)}(q)$ as
\begin{equation}
 \sum_\beta \left(\sum_\alpha L_{\beta\alpha}^{(g)}c_\alpha\right)N_\beta^{(n)}(q)=0,
 \label{eq:linear_system_expanded}
\end{equation}
where $L_{\beta\alpha}^{(g)}$ is a numerical matrix determined by the action of the symmetry operation $g$ on the strain variables.
Because the monomials $N_\beta^{(n)}$ are independent, Eq.~\eqref{eq:linear_system_expanded} gives the linear equations
\begin{equation}
 \sum_\alpha L_{\beta\alpha}^{(g)}c_\alpha=0
 \quad
 \text{for all }\beta\text{ and all generators }g .
 \label{eq:linear_system_coeff}
\end{equation}
Thus, if the coefficient of a monomial is $c_X+c_Y$, the condition is $c_X+c_Y=0$, equivalently $c_X=-c_Y$; it does not require $c_X=c_Y=0$ separately.
By contrast, an identity $c_X X+c_Y Y=0$ for arbitrary independent variables $X$ and $Y$ requires $c_X=0$ and $c_Y=0$.
The allowed strain fields at degree $n$ are the null vectors of the combined linear system, and increasing $n$ from 1 upward gives the leading order and the corresponding basis functions.

As a concrete example, consider the basal field in $4/mmm$.
At first order, write
\begin{equation}
 h_z^{(1)}=c_X X+c_Y Y.
 \label{eq:4mmm_linear_ansatz}
\end{equation}
Under a fourfold rotation about $z$,
\begin{equation}
 (X,Y,Z,U,V)\mapsto(-X,-Y,Z,-V,U),
 \label{eq:C4_action}
\end{equation}
and $A_z^{\mathrm{loc}}$ is even because $C_{4z}$ is a proper rotation, namely $\chi_{A_z^{\mathrm{loc}}}(C_{4z})=+1$.
Equation~\eqref{eq:linear_constraint} therefore gives
\begin{equation}
 -c_X X-c_Y Y=c_X X+c_Y Y.
\end{equation}
Since $X$ and $Y$ are independent variables, this imposes
\begin{equation}
 c_X=0,
 \qquad
 c_Y=0.
\end{equation}
Hence no first-order basal field is allowed.
At second order, write
\begin{equation}
 h_z^{(2)}=c_{X^2}X^2+c_{XY}XY+c_{Y^2}Y^2.
 \label{eq:4mmm_quadratic_ansatz}
\end{equation}
The same $C_{4z}$ operation leaves all three quadratic monomials invariant and gives no constraint.
A vertical mirror, for example the $xz$ mirror, acts as
\begin{equation}
 (X,Y)\mapsto(X,-Y),
 \qquad
 \chi_{A_z^{\mathrm{loc}}}(m_{xz})=-1,
\end{equation}
because $A_z$ is an axial component perpendicular to the mirror plane.
Equation~\eqref{eq:linear_constraint} then gives
\begin{equation}
 c_{X^2}X^2-c_{XY}XY+c_{Y^2}Y^2
 =
 -c_{X^2}X^2-c_{XY}XY-c_{Y^2}Y^2.
\end{equation}
Equating the coefficients of the independent monomials gives
\begin{equation}
 c_{X^2}=0,
 \qquad
 c_{Y^2}=0,
\end{equation}
while $c_{XY}$ remains free.
Therefore the leading basal field in $4/mmm$ is
\begin{equation}
 h_z^{(2)}\propto XY.
\end{equation}

For cubic rows, the field must first be constructed as a vector in the original cubic axes.
We separate the cubic groups into two classes.
For cubic-I groups, $23$ and $m\bar 3$, the lowest-order cubic-covariant axial-vector field is the linear shear triplet
\begin{equation}
\bm h_{\rm I}^{(1)}\propto
\left(
2\varepsilon_{yz}^{\mathrm{cub}},
2\varepsilon_{zx}^{\mathrm{cub}},
2\varepsilon_{xy}^{\mathrm{cub}}
\right).
\label{eq:cubicI_field}
\end{equation}
For cubic-II groups, $432$, $\bar4 3m$, and $m\bar 3m$, no linear strain transforms as an axial vector. A lowest-order cubic-covariant axial-vector field is
\begin{equation}
\bm h_{\rm II}^{(2)}\propto
\begin{pmatrix}
2\varepsilon_{yz}^{\mathrm{cub}}(\varepsilon_{yy}^{\mathrm{cub}}-\varepsilon_{zz}^{\mathrm{cub}})\\
2\varepsilon_{zx}^{\mathrm{cub}}(\varepsilon_{zz}^{\mathrm{cub}}-\varepsilon_{xx}^{\mathrm{cub}})\\
2\varepsilon_{xy}^{\mathrm{cub}}(\varepsilon_{xx}^{\mathrm{cub}}-\varepsilon_{yy}^{\mathrm{cub}})
\end{pmatrix}.
\label{eq:cubicII_field}
\end{equation}
If $C=(\hat{\bm x}_{\mathrm{loc}},\hat{\bm y}_{\mathrm{loc}},\hat{\bm z}_{\mathrm{loc}})$ is the matrix whose columns are the local basis vectors written in cubic axes, then
\begin{equation}
 \bm\varepsilon^{\mathrm{cub}}=C\bm\varepsilon^{\mathrm{loc}}C^{\mathsf T},
 \qquad
 h_{\hat{\bm z}_{\mathrm{loc}}}=\hat{\bm z}_{\mathrm{loc}}\cdot\bm h^{\mathrm{cub}}(\bm\varepsilon^{\mathrm{cub}}).
 \label{eq:cubic_projection}
\end{equation}
The local frames used for the selected cubic axes are listed in Table~\ref{tab:cubic_axes}.

\begin{table}[t]
\caption{Local frames for selected directions in cubic crystals. The columns of $C$ are $(\hat{\bm x}_{\mathrm{loc}},\hat{\bm y}_{\mathrm{loc}},\hat{\bm z}_{\mathrm{loc}})$ expressed in the original cubic axes.}
\label{tab:cubic_axes}
\centering
\small
\begin{tabular}{@{}c c c c@{}}
\toprule
Selected axis & $\hat{\bm x}_{\mathrm{loc}}$ & $\hat{\bm y}_{\mathrm{loc}}$ & $\hat{\bm z}_{\mathrm{loc}}$ \\
\midrule
$[001]$ & $[100]$ & $[010]$ & $[001]$ \\
$[110]$ & $[-110]/\sqrt2$ & $[001]$ & $[110]/\sqrt2$ \\
$[111]$ & $[-110]/\sqrt2$ & $[-1\,-1\,2]/\sqrt6$ & $[111]/\sqrt3$ \\
\bottomrule
\end{tabular}
\end{table}

Equations~\eqref{eq:cubicI_field}--\eqref{eq:cubic_projection} give the cubic local fields summarized in Table~II of the main text.
For example,
\begin{align}
\hat{\bm z}_{[001]}\cdot\bm h_{\rm I}^{(1)}&\propto Y_{\mathrm{loc}},\\
\hat{\bm z}_{[110]}\cdot\bm h_{\rm I}^{(1)}&\propto V_{\mathrm{loc}},\\
\hat{\bm z}_{[111]}\cdot\bm h_{\rm I}^{(1)}&\propto Z_{\mathrm{loc}}.
\end{align}

\section{Derivation of strained subgroups}

The strained subgroup is the stabilizer of the actual strain tensor, not the stabilizer of a scalar coefficient appearing in the free energy.
For a parent point group $G_0$ and a representative strain $q_0$, the residual point group is
\begin{equation}
G_\varepsilon(q_0)=\left\{g\in G_0\mid g\bm\varepsilon(q_0)g^{\mathsf T}=\bm\varepsilon(q_0)\right\}.
\label{eq:stabilizer}
\end{equation}
Equivalently, using Eq.~\eqref{eq:rho_eps_def},
\begin{equation}
G_\varepsilon(q_0)=\left\{g\in G_0\mid \rho_\varepsilon(g)q_0=q_0\right\}.
\label{eq:stabilizer_q}
\end{equation}
The representative $q_0$ is chosen to make the listed leading field nonzero. Except when the row intentionally describes a special high-symmetry strain, $q_0$ is chosen on a generic direction so that no accidental symmetry is introduced.

For the $4/mmm$ basal row, the leading field is $XY$ and a representative strain is
\begin{equation}
q_0=(T,X,Y,Z,U,V)=(0,1,1,0,0,0).
\label{eq:q0_4mmm}
\end{equation}
Although Eq.~\eqref{eq:C4_action} gives $XY\mapsto XY$, it also gives
\begin{equation}
\rho_\varepsilon(C_{4z})q_0=(0,-1,-1,0,0,0)\ne q_0.
\end{equation}
Therefore $C_{4z}$ is not a symmetry of the strained crystal.
The operations satisfying Eq.~\eqref{eq:stabilizer} are
\begin{equation}
 E,
 \quad C_{2z},
 \quad i,
 \quad m_z=iC_{2z},
\end{equation}
so the strained subgroup is $2/m$.

For cubic rows, Eq.~\eqref{eq:stabilizer} is evaluated in the cubic frame. If the representative local strain is $\bm\varepsilon^{\mathrm{loc}}(q_0)$, then
\begin{align}
 \bm\varepsilon^{\mathrm{cub}}_0
 &=C\bm\varepsilon^{\mathrm{loc}}(q_0)C^{\mathsf T},\\
G_\varepsilon^{\mathrm{cub}}(q_0)
&=\left\{g\in G_0^{\mathrm{cub}}\mid
 g\bm\varepsilon^{\mathrm{cub}}_0g^{\mathsf T}=\bm\varepsilon^{\mathrm{cub}}_0\right\}.
\label{eq:cubic_stabilizer}
\end{align}
As an example, for the cubic-I $[111]$ non-basal field $Z_{\mathrm{loc}}$, choose
\begin{equation}
q_0=(0,0,0,1,0,0).
\end{equation}
Using the $[111]$ local frame in Table~\ref{tab:cubic_axes}, Eq.~\eqref{eq:cubic_stabilizer} gives
\begin{equation}
\bm\varepsilon^{\mathrm{cub}}_0=\frac{1}{6}
\begin{pmatrix}
0&1&1\\
1&0&1\\
1&1&0
\end{pmatrix}.
\label{eq:cubic_111_tensor}
\end{equation}
This tensor is invariant under the threefold rotations about $[111]$. Hence the strained subgroup is $3$ for $23$ and $\bar3$ for $m\bar3$.

\end{document}